\documentclass[%
 aip,rsi,amsmath,amssymb,reprint,
]{revtex4-1}

\usepackage{graphicx}
\usepackage{dcolumn}
\usepackage{bm}
\usepackage{tcolorbox}

\begin{document}

\preprint{AIP/123-QED}

\title{Quantum neuromorphic computing}

\author{Danijela Markovi\'c}\thanks{Author to whom correspondance should be addressed. Electronic mail : danijela.markovic@cnrs-thales.fr}
 \affiliation{Unit\'e Mixte de Physique, CNRS, Thales, Universit\'e Paris-Saclay, 91767 Palaiseau, France}
\author{Julie Grollier}
\affiliation{Unit\'e Mixte de Physique, CNRS, Thales, Universit\'e Paris-Saclay, 91767 Palaiseau, France}

\begin{abstract}
Quantum neuromorphic computing physically implements neural networks in brain-inspired quantum hardware to speed up their computation. In this perspective article, we show that this emerging paradigm could make the best use of the existing and near future intermediate size quantum computers. Some approaches are based on parametrized quantum circuits, and use neural network-inspired algorithms to train them. Other approaches, closer to classical neuromorphic computing, take advantage of the physical properties of quantum oscillator assemblies to mimic neurons and compute. We discuss the different implementations of quantum neuromorphic networks with digital and analog circuits, highlight their respective advantages, and review exciting recent experimental results.
\end{abstract}

\maketitle

Over the past decades, quantum and neuromorphic computing have emerged as two leading visions for the future of computation. Quantum computing makes use of intrinsically quantum properties such as entanglement and superposition to design algorithms that are faster than classical ones for some class of problems. On the other hand, neuromorphic computing gets inspiration from the brain and uses complex ensembles of artificial neurons and synapses to mimic animal intelligence and calculate faster with low energy consumption. In this article we review different convergences between these two fields, focusing particularly on the experimental implementations of neuromorphic computing on quantum hardware. We first give a reminder on the two main approaches to quantum computing that are gate-based quantum computing and analog quantum computing. Then we provide an overview of different brain-inspired computing systems, including artificial neural networks that run on general purpose hardware, and neuromorphic networks that run on dedicated hardware. In the core of the article, we review different proposals and experimental implementations of quantum neural networks. We divide them in two groups, digital, implemented on gate-based quantum computers and analog, exploiting the dynamics of quantum annealers and more general disordered quantum systems.

\begin{figure}
    \centering
    \includegraphics[width=1\columnwidth]{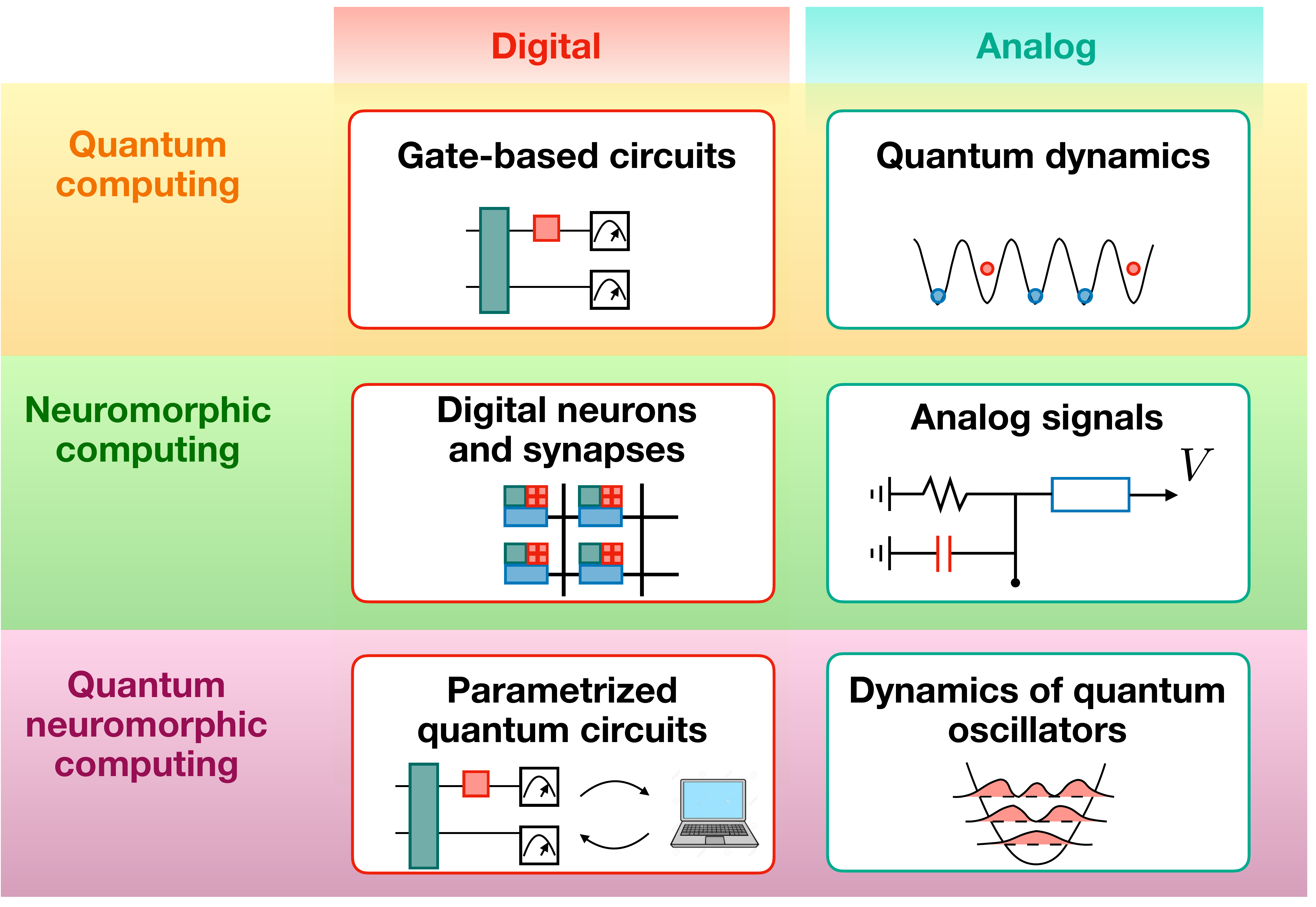}
    \caption{Analog versus digital computing on the different quantum and neuromorphic platforms.}
    \label{digital_analog}
\end{figure}

The two main approaches to quantum computing are digital gate-based quantum computing and analog quantum computing (Fig.~\ref{digital_analog}). \textit{Gate-based quantum computing} uses quantum circuits composed of qubits whose state is manipulated through quantum gates. Quantum gates are reversible unitary operations, such as rotations of a single qubit, or conditional gates that involve two or more qubits and that can be used to entangle them. A gate-based quantum computer is computationally equivalent to a universal quantum computer, meaning that it can express any quantum algorithm.\cite{nielsen_chuang_2010} 
Universal quantum computers are required in order to implement the celebrated quantum algorithms such as Shor's and Grover's algorithm, that could provide respectively exponential and quadratic advantage compared to the best classical algorithms.\cite{Shor1997, Grover1996}
However, today these algorithms cannot beat classical computers even on the biggest existing gate-based quantum computers, commonly called Noisy-Intermediate-Scale-Quantum (NISQ) devices.\cite{Preskill2018} NISQ devices can implement single qubit rotations and an entangling two-qubit gate which form a universal set of quantum gates, meaning that they can express any unitary transformation. On the other hand, the noise in these devices, coming from the interaction with the environment, induces quantum decoherence and destroys the fragile quantum state, thus limiting the maximum depth of circuits that can be realized.

\textit{Analog approaches to quantum computing} include quantum simulation, quantum annealing and adiabatic quantum computing. They rely on quantum dynamics to compute. \textit{Quantum simulators} are quantum systems in which interactions can be controlled such that under the appropriate drives, they emulate specific Hamiltonians.\cite{Lloyd1996} They require a smaller number of operations than gate-based quantum computers and do not need to be completely coherent or without errors, such that they can be realized with available quantum technologies. Quantum simulators allow studying the ground state, temporal dynamics or phase transitions in complex quantum systems. Some examples include open quantum systems simulated using five trapped ion qubits,\cite{Barreiro2011} ultrastrong coupling simulated with simple superconducting circuits \cite{Langford2017, Markovic2018} and a Mott insulator of photons simulated with eight superconducting qubits.\cite{Ma2019}

\textit{Quantum annealers} can implement the transverse Ising spin Hamiltonian on a set of qubits $i$
\begin{equation}
\hat{H}(t) = -\sum_{i,j} J_{ij} \hat{Z}_i \hat{Z}_j - \sum_i h_i \hat{Z}_i - \Gamma(t) \sum_i \hat{X}_i, 
\end{equation}
where $\hat{X}_i$ and $\hat{Z}_i$ are Pauli operators acting on the qubit $i$ and $h_i$ and $J_{ij}$ are adjustable parameters corresponding to local fields and mutual couplings. Quantum annealers are particularly well suited for solving combinatorial optimization problems. The transverse time-dependent field plays the same role as temperature in thermal annealing, leading the system to converge to the lowest energy state.\cite{Kadowaki1998}
The problem is defined by setting the adjustable parameters for each qubit, such that the ground state of the system is the solution of the optimization problem. Compared to classical thermal annealers, they benefit from the phenomenon of quantum tunneling to converge faster to the global minimum and not get stuck in the local minima.\cite{Boixo2014} The availability of stochastic quantum annealers manufactured by D-Wave Systems Inc allowed multiple experimental studies of a subclass of stochastic Hamiltonians whose ground state can be expressed as a classical probability distribution.\cite{Johnson2011}

\textit{Adiabatic quantum computing} is based on adiabatic evolution of a quantum system from the initial state which is the ground state of a simple Hamiltonian, to the complicated Hamiltonian that encodes the problem of interest.\cite{Albash2018} The adiabatic evolution ensures that the system stays in the ground state. Adiabatic quantum computing has been shown to be polynomially equivalent to gate-based quantum computing \cite{Aharonov2008} and thus to be universal, while this is not the case for quantum annealing. It was used on fifty-three trapped ions to simulate a dynamical phase transition. \cite{Zhang2017}

\begin{figure}
    \centering
    \includegraphics[width=1\columnwidth]{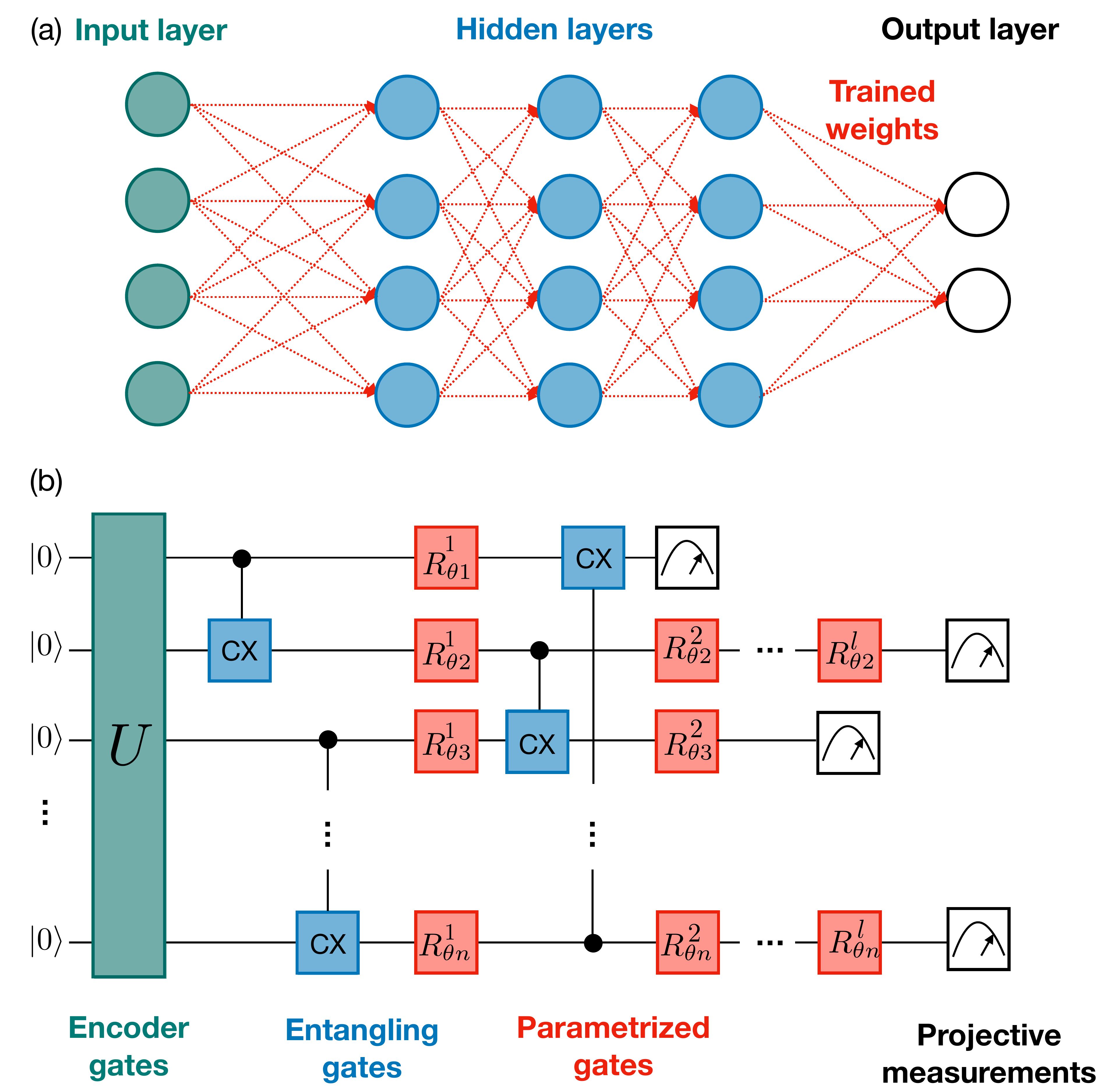}
    \caption{(a) Schematic of an artificial neural network. (b) An example of a parametrized quantum circuit. The encoder circuit $U$ can be decomposed on a set of unitary gates. There are $l$ layers of parametrized gates. Some qubits can be measured in the intermediate layers to reduce the number of degrees of freedom and induce nonlinearity.}
    \label{networks}
\end{figure}

Demonstration of quantum advantage on a useful task on any of the quantum computing platforms, digital or analog, remains elusive and the quest for algorithms that can be run more efficiently on a quantum platform is still open. The recent extraordinary success of machine learning has naturally brought researchers in both quantum information and machine learning to reflect on the possible advantages of merging these two fields.\cite{Biamonte2017} Many exciting questions have emerged, e.g. can quantum hardware provide a speed-up to learning or, are there quantum machine learning approaches that could be robust to noise.\cite{Perdomo-Ortiz2018} The machine learning formalism of particular interest to answer these questions is that of \textit{artificial neural networks}. These computing systems inspired by the brain are composed of computing units called neurons, interconnected by synapses (Fig.~\ref{networks}(a)). Each neuron applies a nonlinear activation function to the weighted sum of the outputs from several other neurons. The nonlinearity is crucial for the ability of the network to approximate any probability distribution. Weights are implemented by synapses and trained using algorithms such as the backpropagation of gradients. Particularly successful are deep neural networks, typically composed of millions of neurons assembled in hundreds of layers. Today they are capable of beating humans in tasks such as image recognition.\cite{Geirhos2017} 

Whereas for quantum computing it was clear since the beginning that dedicated hardware based on completely new physics needed to be developed, neural networks have mainly been implemented on general purpose machines. These machines separate memory and processing units, which causes significant energy consumption for data transfer, a problem commonly called \textit{the von Neumann bottleneck}. Deep neural networks contain billions of parameters that need to be repeatedly calculated and adjusted. Their training becomes extremely slow or even impossible without the use of graphical or tensor processing units that parallelize calculations to some extent but are extremely energy consuming. On the other hand, the brain has a distributed topology that makes it extremely energy efficient. Inspired by the brain, \textit{neuromorphic computing} builds hardware adapted to neural networks. Neuromorphic networks emulate neurons and synapses by physical systems whose outputs are respectively nonlinear and linear functions of their real-valued inputs. The first neuromorphic computers were built in CMOS technology, by assembling transistors into physical units that imitate neurons or synapses.\cite{Markovic2020a} More recent approaches to neuromorphic computing exploit the physics of emerging electronic, optical or spintronic systems to miniaturize neurons and synapses. They typically encode input and output signals in physical quantities such as the amplitude of electrical or spin currents, voltage or light (Fig.~\ref{digital_analog}). Some approaches go a step further and use physical oscillators to reproduce the oscillating dynamics of biological neurons. This allows them to use dynamical phenomena such as synchronization or chaos for learning.\cite{Markovic2020a} An array of four spintronic oscillators was trained to perform a vowel recognition task \cite{Romera2017a} and an optical system of four oscillators could perform recognition of four different letters images.\cite{Feldmann2019}

Neuromorphic computing with physical oscillators has contributed to the development of concepts such as reservoir computing.\cite{Appeltant2011, Paquot2012, Torrejon2017, Markovic2019} Reservoirs are a type of recurrent neural network. Recurrent connections between neurons create loops in which information circulates, thus providing the network with memory and making it particularly well adapted to learning on  time-varying data. Its connections are arbitrary and fixed during learning, and only the weights between the reservoir and the output neurons are trained. This largely simplified learning scheme is one of the reasons reservoirs are among the first network architectures that were implemented on physical neuromorphic hardware. 

Reservoir computing experiments with classical physical oscillators have revealed that the rich dynamics and large state space of the reservoir play an important role in its performance. This makes reservoir particularly interesting for implementation on quantum oscillators that could provide complex quantum dynamics and exponentially large state space, while benefiting from the experience acquired through experiments with classical oscillators.\cite{Appeltant2011, Paquot2012, Torrejon2017} The state of a quantum system of $n$ qubits spans a $2^n$ dimensional vector space, enabling it to store large complex valued vectors and matrices. This property is a key advantage for all quantum neural networks, as it potentially provides them with an exponential increase in memory storage or processing power, and is leveraged in quantum neuromorphic computing.

\textit{Quantum neuromorphic computing} implements neural networks on quantum hardware. Depending on the quantum computing platform, different approaches can be divided in two groups: digital approaches using gate-based quantum computers and analog approaches using analog quantum computing platforms (Fig.~\ref{digital_analog}). Neural networks on gate-based quantum computers are implemented as parametrized quantum circuits.\cite{Benedetti2019a} This formalism is well adapted to NISQ devices, adopting a hybrid quantum/classical approach and lowering the requirements on the number of qubits and coherence time compared to fully quantum procedures. By offloading some subroutines such as data pre- and post-processing to a classical computer, they only keep the core quantum operations on the quantum processor, thus using it to its fullest extent. 

The core of a parametrized quantum circuit is a variational circuit, composed of a series of gates, some fixed and some parametrized and adjustable, that are applied to qubits.\cite{Benedetti2019a} Parameters of adjustable gates are trained using classical routines such as stochastic gradient descent or adaptive momentum estimation. In this way they can find approximate solutions to variational optimization problems. Fixed gates are typically conditional entangling gates such as CNOT (CX), and parameterized gates are typically single qubit rotations, with the rotation angle being the trained parameter. The model output is a function of the expectation values of the observables that can be estimated from the measurements.

Parametrized quantum circuits have been used both for supervised and unsupervised learning tasks. Whereas the variational circuit is common to both of them, there are some differences in the data pre- and post-processing. In \textit{supervised learning} tasks such as data classification, labeled data is given as an input to the neural network, and the network output is trained to match the correct data labels. When parametrized quantum circuits are used for classical tasks, classical data first needs to be encoded in the quantum state of the system. This is done by an encoder circuit that precedes the variational circuit (Fig.~\ref{networks}(b)). Encoding methods can be grouped in two main categories, amplitude and qubit encoding.
 
With \textit{amplitude encoding}, inputs are encoded in the probability amplitudes $i_n$ of the basis states $|n\rangle$, by preparing the whole system in the state 
\begin{equation}
|\psi \rangle = \sum_{n=1}^{2^N} i_n |n\rangle,
\label{amplitude_encoding}
\end{equation}
where $|n \rangle \in \{|00...0 \rangle, |00...1 \rangle, ... |11...1 \rangle \}$. This method encodes $2^N$ inputs in $N$ qubits, providing an exponential advantage in terms of storage.  With amplitude encoding unitary gates transform data linearly, thus requiring indirect techniques to introduce the crucial nonlinearity of the neural activation function. It was experimentally demonstrated that  nonlinearity provided by the measurement is sufficient to successfully perform classification tasks using a four neuron quantum perceptron on two superconducting qubits.\cite{Tacchino2019} Unfortunately, the preparation of the superposition state requires a complicated encoder circuit whose depth is expected to scale exponentially with the number of qubits.\cite{Benedetti2019a} This aspect might obstruct the use of amplitude encoding on bigger quantum computers. 

With \textit{qubit encoding}, each input is encoded in the amplitude of a single qubit, such that the $n^{\textrm{th}}$ qubit is prepared in the state
\begin{equation}
|\psi_n\rangle = \cos(i_n) |0\rangle +  \sin(i_n) |1\rangle
\label{qubit_encoding}
\end{equation}
and the state of the whole quantum register is a dis-entangled product state
\begin{equation}
    |\psi\rangle = \otimes_{n=1}^N |\psi_n\rangle.
\end{equation}
Qubit encoding is less space-efficient as it requires $N$ qubits to encode $N$ input values but it is more time and energy efficient as only single qubit rotations are required to realize it. Most importantly, the tensor product produces a number of nonlinear basis functions exponential in the number of qubits \cite{Stoudenmire2016}. This could be a source of advantage with respect to classical approaches where the network depth needs to be increased in order to obtain a larger number of nonlinear basis functions and increase the precision of the model prediction.\cite{Mitarai2018}

Due to the small number of qubits and limited connectivity of the NISQ devices, experimental realizations of supervised learning with parametrized quantum circuits have used very simple or scaled-down datasets.\cite{Grant2018, Havlicek2019, Tacchino2019} In order to compare fairly to classical neural networks, demonstrations on larger number of qubits are needed. Nevertheless, potential sources of advantage provided by quantum hardware, such as quantum-enhanced feature space, were put forward. Controllable entanglement and interference allow for efficient kernel estimation in the exponential feature space.\cite{Havlicek2019} Features produced by a quantum convolutional neural network could increase the accuracy and make the training faster compared to classical convolutional neural networks.\cite{Henderson2020}

In \textit{unsupervised learning} tasks such as clustering or modeling, the neural network looks for patterns in the previously unlabeled data. Most of the unsupervised tasks experimentally performed with parametrized quantum circuits fall in the category of generative modeling.\cite{Benedetti2019a} Generative modeling aims at learning the probability distribution of the training data with the goal of generating new data points. Quantum generative models make use of the probabilistic nature of the quantum system wavefunction, which encodes the generated model distribution.\cite{Han2018} The computational complexity of quantum systems suggests that they should be more expressive than classical models. This type of quantum generative models are implicit, meaning that one can easily sample from the model distribution by performing projective measurements on the qubits. However, the model distribution itself is intractable. Implicit models can be more expressive than the explicit ones, but they are harder to train, because it is more difficult to define a loss function when the model distribution is unknown.\cite{Mohamed2016} Compared to supervised learning problems, the unsupervised learning has the advantage to circumvent the difficulty of data encoding. The parametrized circuit acts on qubits initialized in state $|0\rangle$ and is trained to prepare the wavefunction that encodes the model distribution. The training is done on a classical computer by minimizing the divergence between the model and target distributions. Most experimental realizations of quantum generative modeling used the \textit{Bars and Stripes} dataset of 2 $\times$ 2 black and white images. As it can be modeled with 4 qubits, it is used to benchmark the power and usefulness of NISQ devices. It was used for instance on the IBM quantum chip to analyze the impact of the noise that accumulates with layers of gates, demonstrating that sparsely connected shallow circuits outperform denser ones on noisy hardware.\cite{Hamilton2019} Implementation on a trapped ions system allowed studying the performance dependence on the connectivity topology and number of layers on the same device due to its full connectivity between qubits and demonstrated the key role of the entangling gates.\cite{Benedetti2019b} A similar study was realized on the Rigetti quantum computing platform.\cite{Leyton-Ortega2019}
Probabilistic tasks like generative modeling make use of the capability of quantum computers to efficiently prepare and sample certain probability distributions which makes them the most likely to prove quantum advantage in the near-term.\cite{Perdomo-Ortiz2018}

Quantum neural networks open the possibility to perform inherently quantum tasks such as \textit{classification or modeling of quantum states}. A quantum state can be prepared in a physical system without direct access to its classical description. This procedure thus bypasses costly quantum state tomography which requires an exponential number of copies of the quantum state to be prepared and measured. It was experimentally shown on an optical quantum computer that it is possible to approximately learn a quantum state from a training set containing a linear number of quantum state examples.\cite{Rocchetto2019} Proof-of-principle quantum state classification was performed by preparing two different classes of quantum states by two circuits with different depths and thus different levels of entanglement. They could be classified by an eight superconducting qubit circuit.\cite{Grant2018} Numerical simulations showed that a quantum convolutional neural network implemented on fifteen qubits can realize quantum phase recognition \cite{Cong2019} and that shallow quantum circuits can classify pure and mixed quantum states.\cite{Chen2018a} 

Another successful approach to quantum state learning uses quantum generative adversarial networks (quantum GANs).\cite{Lloyd2018} A GAN in classical machine learning is composed of two networks, a generator and a discriminator, that are trained through an adversarial procedure. The generator optimizes the model parameters to produce fake data, while the discriminator optimizes its parameters to distinguish between the model and the true data. Its quantum version has been shown to have the potential of exhibiting an exponential advantage over the classical one on very high-dimensional datasets.\cite{Lloyd2018} Learning of a single qubit state using a quantum GAN was experimentally demonstrated on a simple superconducting circuit.\cite{Hu2019} Parametrized quantum circuits can also be used to solve classically intractable problems e.g. calculate molecular ground-state energies. A superconducting circuit learned the energy of the H$_2$ molecule with better robustness to noise than gate-based algorithms.\cite{OMalley2016} Trapped ions system with three qubits could learn the energy of a more complicated LiH molecule.\cite{Hempel2018} Finally, they are promising for other quantum learning tasks such as \textit{quantum circuit learning}, which could assist quantum computers by finding optimal circuits for quantum algorithms such as Grover's.\cite{Morales2018}

\textit{Analog quantum neuromorphic computing} relies on the dynamics of a quantum system. It encompasses adiabatic dynamics of quantum annealers and more general dynamics of disordered quantum systems that can be leveraged in the context of reservoir computing, or computing with continuous variables. Experimental implementations of neural networks on quantum annealers have been slowed down because of the limited inter-connectivity in the existing systems, which severely restricts the problem embedding. A solution has been found recently using graphical models embedded in hardware, thus enabling an implementation of unsupervised learning of binarized OptDigits dataset on the D-Wave \textit{quantum annealer}.\cite{Benedetti2017} Under some constraints, quantum annealers could implement quantum Boltzmann machines. Quantum Boltzmann machines  were numerically demonstrated to learn the data distribution of small size examples better than classical Boltzmann machines,\cite{Amin2018} but experimental demonstrations are still lacking. 

\begin{figure}
    \centering
    \includegraphics[width=1\columnwidth]{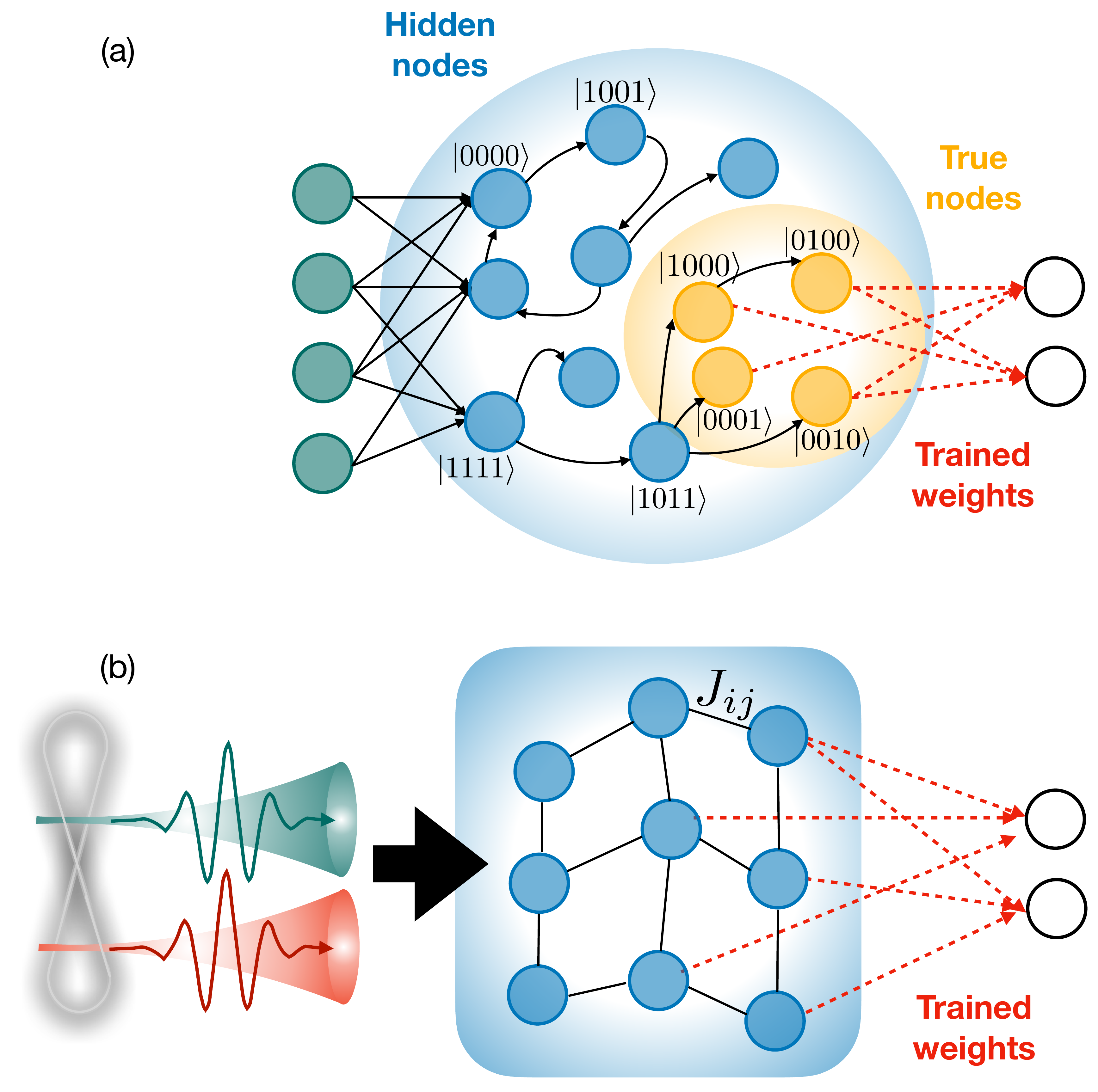}
    \caption{(a) Quantum reservoir using amplitude encoding.\cite{Nakajima2019} All the basis states are reservoir nodes that participate to computation and only the states with a single excitation are monitored. (b) Quantum reservoir based on coupled quantum dots for classification of entangled and separable quantum states. \cite{Ghosh2019}}
    \label{reservoirs}
\end{figure}

 More diverse dynamics of disordered quantum systems can be leveraged for computing by extending the \textit{reservoir} neuromorphic network to quantum domain. Several implementations of a quantum reservoir have been proposed, spotlighting different advantages that quantum hardware brings. First, by using the \textit{amplitude encoding} as in Eq.~\eqref{amplitude_encoding}, the size of the reservoir can be exponentially increased.\cite{Fujii2017} In that scheme, reservoir sites are not qubits but basis states $|n\rangle$ and only a subset of these states, called the true nodes, is measured to determine neural outputs (Fig.~\ref{reservoirs}(a)). All the other non-monitored states are hidden nodes that participate in information processing. Quantum reservoir with amplitude encoding does not suffer from the costly data encoding encountered with parametrized quantum circuits as it uses the natural system dynamics to populate states. Such a reservoir can process classical tasks that require memory, such as the timer task, which consists in giving the output 1 only after the number of time steps given by the input. Compared to a classical Echo State Network (ESN), such quantum reservoir demonstrates an advantage in terms of size. Its performance can be boosted through spatial multiplexing,\cite{Nakajima2019} which is of interest for implementations on existing quantum machines where quantum coherence is not preserved over large distances. This type of reservoir is well adapted for disordered ensemble quantum systems, such as nuclear magnetic resonance or quantum dots, but its realization on quantum gate-based circuits remains challenging due to the large number of gates required to implement the reservoir dynamics. 

 A different form of quantum reservoir with natural quantum circuit interpretation has been proposed.\cite{Chen2020} The input $i_n$ is encoded as the probability to apply a unitary gate $U_0$ or $U_1$, such that the density matrix that describes the system after the input $i_n$ is
 \begin{equation}
     \rho_n = (1-\epsilon)(i_n U_0\rho_{n-1} U_0^\dagger + (1-i_n) U_1\rho_{n-1} U_1^\dagger) + \epsilon \sigma,
 \end{equation}
where $\sigma$ is an arbitrary and fixed density matrix and $\epsilon$ controls the speed at which the reservoir forgets (no memory for $\epsilon=1$). Such reservoir was implemented on several different IBM's superconducting quantum computers, with different numbers of qubits, successfully solving classical tasks of time series prediction \cite{Chen2020} and experimentally demonstrating the spatial multiplexing \cite{Nakajima2019}. For such a reservoir there is an optimal inter-connectivity, which provides nonlinearity but also induces noise and decoherence.\cite{Dasgupta2020}

Quantum reservoirs discussed up to this point focused on classical tasks. Yet, a reservoir implemented on a disordered quantum system can make use of its particular physical properties, which is beneficial to learn inherently quantum tasks. For instance, a reservoir of quantum dots couples to light, such that it can naturally receive squeezed quantum states of light as inputs (Fig.~\ref{reservoirs}(b)). A numerical simulation demonstrated that a reservoir implemented on a disordered quantum dot system can classify entangled and separable quantum states of two light modes.\cite{Ghosh2019} A quantum reservoir can thus act as a quantum detector. Furthermore, it can be trained to prepare different quantum states,\cite{Ghosh2019b} and to perform a universal set of quantum gates between the output qubits.\cite{Ghosh2020}

Another approach to analog quantum neuromorphic computing makes use of continuous quantum variables such as amplitude of the electromagnetic field, instead of discrete valued energy states of qubits, $|0\rangle$ and $|1\rangle$. This type of computing can be implemented on photonic or microwave circuits. The quantum state can be manipulated through rotation, displacement and squeezing operations and nonlinearity can be provided by the Kerr effect. Continuous variable quantum neural network was used in numerical simulations to solve classical tasks such as nonlinear function fitting and quantum tasks such as learning the phase space encoding of quantum Fock states with an autoencoder circuit.\cite{Killoran2018} A kernel-based classifier was trained using a continuous variable quantum circuit.\cite{Schuld2019} A continuous variable reservoir using a single quantum oscillator was numerically simulated and demonstrated improvement compared to a classical reservoir on the classical task of sine phase estimation.\cite{Govia2020} Continuous variable computing is suggested to be more robust to errors and to reduce the number of measurements needed to estimate the expectation values of quantum variables.  
Reservoir performance was shown to be better for a noisy open quantum system that exchanges photons with the environment, as it prevents overfitting.

In conclusion, both deep neural networks and quantum computing have been developing very quickly in the past years with more and more researchers entering the fields. The merging of these two fields was approached from various sides by researchers with different backgrounds. We attempted to give a framework to interpret these results by discussing them from the point of view of quantum neuromorphic computing. Quantum neuromorphic computing is now challenged to answer many foundational questions. Is quantum neural network learning more robust to noise than other gate-based quantum procedures?
Could quantum hardware provide a speed up to neuromorphic computing? Can quantum neuromorphic computing solve the von Neumann bottleneck? 

Preliminary answers have been put forward. There are some indications that quantum neural networks are more robust to noise than gate-based algorithms. For quantum reservoir computing, decoherence to some degree is even beneficial as it improves nonlinearity, though in the large decoherence limit, the system becomes classical. Quantum advantage is provided by the exponentially large state space of coupled quantum systems, but it is compromised when encoding classical data in the state of a quantum system and when reconstructing the classical representation of a quantum state, as both of these procedures require an exponential number of operations. A quantum speed-up is thus likely to be obtained with algorithms that circumvent this "quantum-to and from-classical  bottleneck". Among these are implicit unsupervised learning on classical data as well as supervised and unsupervised learning on quantum data. Finally, learning on quantum data is particularly promising with the emerging analog quantum neural networks, which make use of the physics of the particular quantum system to directly couple to input quantum states provided by another system, for example through light-matter coupling.

Work by D.M. on neuromorphic computing with oscillators was supported by the European Research Council ERC under bioSPINspired Grant No. 682955. J.G acknowledges support by DOE BES Award \# DE-SC0019273 on quantum neuromorphic computing for Q-MEEN-C, an Energy Frontier Research Center funded by the U.S. Department of Energy (DOE), Office of Science, Basic Energy Sciences (BES).
 We would like to thank Mark Stiles, Andrew Kent and Axel Hoffmann for fruitful discussions within the Q-MEEN-C EFRC.


\begin{thebibliography}{57}%
\makeatletter
\providecommand \@ifxundefined [1]{%
 \@ifx{#1\undefined}
}%
\providecommand \@ifnum [1]{%
 \ifnum #1\expandafter \@firstoftwo
 \else \expandafter \@secondoftwo
 \fi
}%
\providecommand \@ifx [1]{%
 \ifx #1\expandafter \@firstoftwo
 \else \expandafter \@secondoftwo
 \fi
}%
\providecommand \natexlab [1]{#1}%
\providecommand \enquote  [1]{``#1''}%
\providecommand \bibnamefont  [1]{#1}%
\providecommand \bibfnamefont [1]{#1}%
\providecommand \citenamefont [1]{#1}%
\providecommand \href@noop [0]{\@secondoftwo}%
\providecommand \href [0]{\begingroup \@sanitize@url \@href}%
\providecommand \@href[1]{\@@startlink{#1}\@@href}%
\providecommand \@@href[1]{\endgroup#1\@@endlink}%
\providecommand \@sanitize@url [0]{\catcode `\\12\catcode `\$12\catcode
  `\&12\catcode `\#12\catcode `\^12\catcode `\_12\catcode `\%12\relax}%
\providecommand \@@startlink[1]{}%
\providecommand \@@endlink[0]{}%
\providecommand \url  [0]{\begingroup\@sanitize@url \@url }%
\providecommand \@url [1]{\endgroup\@href {#1}{\urlprefix }}%
\providecommand \urlprefix  [0]{URL }%
\providecommand \Eprint [0]{\href }%
\providecommand \doibase [0]{http://dx.doi.org/}%
\providecommand \selectlanguage [0]{\@gobble}%
\providecommand \bibinfo  [0]{\@secondoftwo}%
\providecommand \bibfield  [0]{\@secondoftwo}%
\providecommand \translation [1]{[#1]}%
\providecommand \BibitemOpen [0]{}%
\providecommand \bibitemStop [0]{}%
\providecommand \bibitemNoStop [0]{.\EOS\space}%
\providecommand \EOS [0]{\spacefactor3000\relax}%
\providecommand \BibitemShut  [1]{\csname bibitem#1\endcsname}%
\let\auto@bib@innerbib\@empty
\bibitem [{\citenamefont {Nielsen}\ and\ \citenamefont
  {Chuang}(2010)}]{nielsen_chuang_2010}%
  \BibitemOpen
  \bibfield  {author} {\bibinfo {author} {\bibfnamefont {M.~A.}\ \bibnamefont
  {Nielsen}}\ and\ \bibinfo {author} {\bibfnamefont {I.~L.}\ \bibnamefont
  {Chuang}},\ }\href {\doibase 10.1017/CBO9780511976667} {\emph {\bibinfo
  {title} {Quantum Computation and Quantum Information: 10th Anniversary
  Edition}}}\ (\bibinfo  {publisher} {Cambridge University Press},\ \bibinfo
  {year} {2010})\BibitemShut {NoStop}%
\bibitem [{\citenamefont {Shor}(1997)}]{Shor1997}%
  \BibitemOpen
  \bibfield  {author} {\bibinfo {author} {\bibfnamefont {P.}~\bibnamefont
  {Shor}},\ }\href@noop {} {\bibfield  {journal} {\bibinfo  {journal} {SIAM
  Journal on Computing}\ }\textbf {\bibinfo {volume} {26}},\ \bibinfo {pages}
  {1484} (\bibinfo {year} {1997})}\BibitemShut {NoStop}%
\bibitem [{\citenamefont {Grover}(1996)}]{Grover1996}%
  \BibitemOpen
  \bibfield  {author} {\bibinfo {author} {\bibfnamefont {L.}~\bibnamefont
  {Grover}},\ }\href@noop {} {\bibfield  {journal} {\bibinfo  {journal} {28th
  Annual ACM Symposium on the Theory of Computing}\ ,\ \bibinfo {pages} {212}}
  (\bibinfo {year} {1996})}\BibitemShut {NoStop}%
\bibitem [{\citenamefont {Preskill}(2018)}]{Preskill2018}%
  \BibitemOpen
  \bibfield  {author} {\bibinfo {author} {\bibfnamefont {J.}~\bibnamefont
  {Preskill}},\ }\href@noop {} {\bibfield  {journal} {\bibinfo  {journal}
  {arXiv 1801.00862}\ } (\bibinfo {year} {2018})}\BibitemShut {NoStop}%
\bibitem [{\citenamefont {Lloyd}(1996)}]{Lloyd1996}%
  \BibitemOpen
  \bibfield  {author} {\bibinfo {author} {\bibfnamefont {S.}~\bibnamefont
  {Lloyd}},\ }\href@noop {} {\bibfield  {journal} {\bibinfo  {journal}
  {Science}\ }\textbf {\bibinfo {volume} {273}},\ \bibinfo {pages} {1073}
  (\bibinfo {year} {1996})}\BibitemShut {NoStop}%
\bibitem [{\citenamefont {Barreiro}\ \emph {et~al.}(2011)\citenamefont
  {Barreiro}, \citenamefont {M{\"{u}}ller}, \citenamefont {Schindler},
  \citenamefont {Nigg}, \citenamefont {Monz}, \citenamefont {Chwalla},
  \citenamefont {Hennrich}, \citenamefont {Roos}, \citenamefont {Zoller},\ and\
  \citenamefont {Blatt}}]{Barreiro2011}%
  \BibitemOpen
  \bibfield  {author} {\bibinfo {author} {\bibfnamefont {J.~T.}\ \bibnamefont
  {Barreiro}}, \bibinfo {author} {\bibfnamefont {M.}~\bibnamefont
  {M{\"{u}}ller}}, \bibinfo {author} {\bibfnamefont {P.}~\bibnamefont
  {Schindler}}, \bibinfo {author} {\bibfnamefont {D.}~\bibnamefont {Nigg}},
  \bibinfo {author} {\bibfnamefont {T.}~\bibnamefont {Monz}}, \bibinfo {author}
  {\bibfnamefont {M.}~\bibnamefont {Chwalla}}, \bibinfo {author} {\bibfnamefont
  {M.}~\bibnamefont {Hennrich}}, \bibinfo {author} {\bibfnamefont {C.~F.}\
  \bibnamefont {Roos}}, \bibinfo {author} {\bibfnamefont {P.}~\bibnamefont
  {Zoller}}, \ and\ \bibinfo {author} {\bibfnamefont {R.}~\bibnamefont
  {Blatt}},\ }\href {\doibase 10.1038/nature09801} {\bibfield  {journal}
  {\bibinfo  {journal} {Nature}\ }\textbf {\bibinfo {volume} {470}},\ \bibinfo
  {pages} {486} (\bibinfo {year} {2011})}\BibitemShut {NoStop}%
\bibitem [{\citenamefont {Langford}\ \emph {et~al.}(2017)\citenamefont
  {Langford}, \citenamefont {Sagastizabal}, \citenamefont {Kounalakis},
  \citenamefont {Dickel}, \citenamefont {Bruno}, \citenamefont {Luthi},
  \citenamefont {Thoen}, \citenamefont {Endo},\ and\ \citenamefont
  {Dicarlo}}]{Langford2017}%
  \BibitemOpen
  \bibfield  {author} {\bibinfo {author} {\bibfnamefont {N.~K.}\ \bibnamefont
  {Langford}}, \bibinfo {author} {\bibfnamefont {R.}~\bibnamefont
  {Sagastizabal}}, \bibinfo {author} {\bibfnamefont {M.}~\bibnamefont
  {Kounalakis}}, \bibinfo {author} {\bibfnamefont {C.}~\bibnamefont {Dickel}},
  \bibinfo {author} {\bibfnamefont {A.}~\bibnamefont {Bruno}}, \bibinfo
  {author} {\bibfnamefont {F.}~\bibnamefont {Luthi}}, \bibinfo {author}
  {\bibfnamefont {D.~J.}\ \bibnamefont {Thoen}}, \bibinfo {author}
  {\bibfnamefont {A.}~\bibnamefont {Endo}}, \ and\ \bibinfo {author}
  {\bibfnamefont {L.}~\bibnamefont {Dicarlo}},\ }\href@noop {} {\bibfield
  {journal} {\bibinfo  {journal} {Nature Communications}\ }\textbf {\bibinfo
  {volume} {8}} (\bibinfo {year} {2017})}\BibitemShut {NoStop}%
\bibitem [{\citenamefont {Markovi{\'{c}}}\ \emph {et~al.}(2018)\citenamefont
  {Markovi{\'{c}}}, \citenamefont {Jezouin}, \citenamefont {Ficheux},
  \citenamefont {Fedortchenko}, \citenamefont {Felicetti}, \citenamefont
  {Coudreau}, \citenamefont {Milman}, \citenamefont {Leghtas},\ and\
  \citenamefont {Huard}}]{Markovic2018}%
  \BibitemOpen
  \bibfield  {author} {\bibinfo {author} {\bibfnamefont {D.}~\bibnamefont
  {Markovi{\'{c}}}}, \bibinfo {author} {\bibfnamefont {S.}~\bibnamefont
  {Jezouin}}, \bibinfo {author} {\bibfnamefont {Q.}~\bibnamefont {Ficheux}},
  \bibinfo {author} {\bibfnamefont {S.}~\bibnamefont {Fedortchenko}}, \bibinfo
  {author} {\bibfnamefont {S.}~\bibnamefont {Felicetti}}, \bibinfo {author}
  {\bibfnamefont {T.}~\bibnamefont {Coudreau}}, \bibinfo {author}
  {\bibfnamefont {P.}~\bibnamefont {Milman}}, \bibinfo {author} {\bibfnamefont
  {Z.}~\bibnamefont {Leghtas}}, \ and\ \bibinfo {author} {\bibfnamefont
  {B.}~\bibnamefont {Huard}},\ }\href@noop {} {\bibfield  {journal} {\bibinfo
  {journal} {Physical Review Letters}\ }\textbf {\bibinfo {volume} {121}}
  (\bibinfo {year} {2018})}\BibitemShut {NoStop}%
\bibitem [{\citenamefont {Ma}\ \emph {et~al.}(2019)\citenamefont {Ma},
  \citenamefont {Saxberg}, \citenamefont {Owens}, \citenamefont {Leung},
  \citenamefont {Lu}, \citenamefont {Simon},\ and\ \citenamefont
  {Schuster}}]{Ma2019}%
  \BibitemOpen
  \bibfield  {author} {\bibinfo {author} {\bibfnamefont {R.}~\bibnamefont
  {Ma}}, \bibinfo {author} {\bibfnamefont {B.}~\bibnamefont {Saxberg}},
  \bibinfo {author} {\bibfnamefont {C.}~\bibnamefont {Owens}}, \bibinfo
  {author} {\bibfnamefont {N.}~\bibnamefont {Leung}}, \bibinfo {author}
  {\bibfnamefont {Y.}~\bibnamefont {Lu}}, \bibinfo {author} {\bibfnamefont
  {J.}~\bibnamefont {Simon}}, \ and\ \bibinfo {author} {\bibfnamefont {D.~I.}\
  \bibnamefont {Schuster}},\ }\href@noop {} {\bibfield  {journal} {\bibinfo
  {journal} {Nature}\ }\textbf {\bibinfo {volume} {566}},\ \bibinfo {pages}
  {51} (\bibinfo {year} {2019})}\BibitemShut {NoStop}%
\bibitem [{\citenamefont {Kadowaki}\ and\ \citenamefont
  {Nishimori}(1998)}]{Kadowaki1998}%
  \BibitemOpen
  \bibfield  {author} {\bibinfo {author} {\bibfnamefont {T.}~\bibnamefont
  {Kadowaki}}\ and\ \bibinfo {author} {\bibfnamefont {H.}~\bibnamefont
  {Nishimori}},\ }\href {\doibase 10.1103/PhysRevE.58.5355} {\bibfield
  {journal} {\bibinfo  {journal} {Physical Review E}\ }\textbf {\bibinfo
  {volume} {58}},\ \bibinfo {pages} {5} (\bibinfo {year} {1998})}\BibitemShut
  {NoStop}%
\bibitem [{\citenamefont {Boixo}\ \emph {et~al.}(2014)\citenamefont {Boixo},
  \citenamefont {R{\o}nnow}, \citenamefont {Isakov}, \citenamefont {Wang},
  \citenamefont {Wecker}, \citenamefont {Lidar}, \citenamefont {Martinis},\
  and\ \citenamefont {Troyer}}]{Boixo2014}%
  \BibitemOpen
  \bibfield  {author} {\bibinfo {author} {\bibfnamefont {S.}~\bibnamefont
  {Boixo}}, \bibinfo {author} {\bibfnamefont {T.~F.}\ \bibnamefont
  {R{\o}nnow}}, \bibinfo {author} {\bibfnamefont {S.~V.}\ \bibnamefont
  {Isakov}}, \bibinfo {author} {\bibfnamefont {Z.}~\bibnamefont {Wang}},
  \bibinfo {author} {\bibfnamefont {D.}~\bibnamefont {Wecker}}, \bibinfo
  {author} {\bibfnamefont {D.~A.}\ \bibnamefont {Lidar}}, \bibinfo {author}
  {\bibfnamefont {J.~M.}\ \bibnamefont {Martinis}}, \ and\ \bibinfo {author}
  {\bibfnamefont {M.}~\bibnamefont {Troyer}},\ }\href {\doibase
  10.1038/nphys2900} {\bibfield  {journal} {\bibinfo  {journal} {Nature
  Physics}\ }\textbf {\bibinfo {volume} {10}},\ \bibinfo {pages} {218}
  (\bibinfo {year} {2014})}\BibitemShut {NoStop}%
\bibitem [{\citenamefont {Johnson}\ \emph {et~al.}(2011)\citenamefont
  {Johnson}, \citenamefont {Amin}, \citenamefont {Gildert}, \citenamefont
  {Lanting}, \citenamefont {Hamze}, \citenamefont {Dickson}, \citenamefont
  {Harris}, \citenamefont {Berkley}, \citenamefont {Johansson}, \citenamefont
  {Bunyk}, \citenamefont {Chapple}, \citenamefont {Enderud}, \citenamefont
  {Hilton}, \citenamefont {Karimi}, \citenamefont {Ladizinsky}, \citenamefont
  {Ladizinsky}, \citenamefont {Oh}, \citenamefont {Perminov}, \citenamefont
  {Rich}, \citenamefont {Thom}, \citenamefont {Tolkacheva}, \citenamefont
  {Truncik}, \citenamefont {Uchaikin}, \citenamefont {Wang}, \citenamefont
  {Wilson},\ and\ \citenamefont {Rose}}]{Johnson2011}%
  \BibitemOpen
  \bibfield  {author} {\bibinfo {author} {\bibfnamefont {M.~W.}\ \bibnamefont
  {Johnson}}, \bibinfo {author} {\bibfnamefont {M.~H.}\ \bibnamefont {Amin}},
  \bibinfo {author} {\bibfnamefont {S.}~\bibnamefont {Gildert}}, \bibinfo
  {author} {\bibfnamefont {T.}~\bibnamefont {Lanting}}, \bibinfo {author}
  {\bibfnamefont {F.}~\bibnamefont {Hamze}}, \bibinfo {author} {\bibfnamefont
  {N.}~\bibnamefont {Dickson}}, \bibinfo {author} {\bibfnamefont
  {R.}~\bibnamefont {Harris}}, \bibinfo {author} {\bibfnamefont {A.~J.}\
  \bibnamefont {Berkley}}, \bibinfo {author} {\bibfnamefont {J.}~\bibnamefont
  {Johansson}}, \bibinfo {author} {\bibfnamefont {P.}~\bibnamefont {Bunyk}},
  \bibinfo {author} {\bibfnamefont {E.~M.}\ \bibnamefont {Chapple}}, \bibinfo
  {author} {\bibfnamefont {C.}~\bibnamefont {Enderud}}, \bibinfo {author}
  {\bibfnamefont {J.~P.}\ \bibnamefont {Hilton}}, \bibinfo {author}
  {\bibfnamefont {K.}~\bibnamefont {Karimi}}, \bibinfo {author} {\bibfnamefont
  {E.}~\bibnamefont {Ladizinsky}}, \bibinfo {author} {\bibfnamefont
  {N.}~\bibnamefont {Ladizinsky}}, \bibinfo {author} {\bibfnamefont
  {T.}~\bibnamefont {Oh}}, \bibinfo {author} {\bibfnamefont {I.}~\bibnamefont
  {Perminov}}, \bibinfo {author} {\bibfnamefont {C.}~\bibnamefont {Rich}},
  \bibinfo {author} {\bibfnamefont {M.~C.}\ \bibnamefont {Thom}}, \bibinfo
  {author} {\bibfnamefont {E.}~\bibnamefont {Tolkacheva}}, \bibinfo {author}
  {\bibfnamefont {C.~J.}\ \bibnamefont {Truncik}}, \bibinfo {author}
  {\bibfnamefont {S.}~\bibnamefont {Uchaikin}}, \bibinfo {author}
  {\bibfnamefont {J.}~\bibnamefont {Wang}}, \bibinfo {author} {\bibfnamefont
  {B.}~\bibnamefont {Wilson}}, \ and\ \bibinfo {author} {\bibfnamefont
  {G.}~\bibnamefont {Rose}},\ }\href {\doibase 10.1038/nature10012} {\bibfield
  {journal} {\bibinfo  {journal} {Nature}\ }\textbf {\bibinfo {volume} {473}},\
  \bibinfo {pages} {194} (\bibinfo {year} {2011})}\BibitemShut {NoStop}%
\bibitem [{\citenamefont {Albash}\ and\ \citenamefont
  {Lidar}(2018)}]{Albash2018}%
  \BibitemOpen
  \bibfield  {author} {\bibinfo {author} {\bibfnamefont {T.}~\bibnamefont
  {Albash}}\ and\ \bibinfo {author} {\bibfnamefont {D.~A.}\ \bibnamefont
  {Lidar}},\ }\href {\doibase 10.1103/RevModPhys.90.015002} {\bibfield
  {journal} {\bibinfo  {journal} {Reviews of Modern Physics}\ }\textbf
  {\bibinfo {volume} {90}},\ \bibinfo {pages} {15002} (\bibinfo {year}
  {2018})}\BibitemShut {NoStop}%
\bibitem [{\citenamefont {Aharonov}\ \emph {et~al.}(2008)\citenamefont
  {Aharonov}, \citenamefont {{Van Dam}}, \citenamefont {Kempe}, \citenamefont
  {Landau}, \citenamefont {Lloyd},\ and\ \citenamefont {Regev}}]{Aharonov2008}%
  \BibitemOpen
  \bibfield  {author} {\bibinfo {author} {\bibfnamefont {D.}~\bibnamefont
  {Aharonov}}, \bibinfo {author} {\bibfnamefont {W.}~\bibnamefont {{Van Dam}}},
  \bibinfo {author} {\bibfnamefont {J.}~\bibnamefont {Kempe}}, \bibinfo
  {author} {\bibfnamefont {Z.}~\bibnamefont {Landau}}, \bibinfo {author}
  {\bibfnamefont {S.}~\bibnamefont {Lloyd}}, \ and\ \bibinfo {author}
  {\bibfnamefont {O.}~\bibnamefont {Regev}},\ }\href {\doibase
  10.1137/080734479} {\bibfield  {journal} {\bibinfo  {journal} {SIAM Jour.
  Comput.}\ }\textbf {\bibinfo {volume} {50}},\ \bibinfo {pages} {755}
  (\bibinfo {year} {2008})}\BibitemShut {NoStop}%
\bibitem [{\citenamefont {Zhang}\ \emph {et~al.}(2017)\citenamefont {Zhang},
  \citenamefont {hyagano}, \citenamefont {Hess}, \citenamefont {Kyprianidis},
  \citenamefont {Becker}, \citenamefont {Kaplan}, \citenamefont {Gorshkov},
  \citenamefont {Gong},\ and\ \citenamefont {Monroe}}]{Zhang2017}%
  \BibitemOpen
  \bibfield  {author} {\bibinfo {author} {\bibfnamefont {J.}~\bibnamefont
  {Zhang}}, \bibinfo {author} {\bibfnamefont {G.}~\bibnamefont {hyagano}},
  \bibinfo {author} {\bibfnamefont {P.~W.}\ \bibnamefont {Hess}}, \bibinfo
  {author} {\bibfnamefont {A.}~\bibnamefont {Kyprianidis}}, \bibinfo {author}
  {\bibfnamefont {P.}~\bibnamefont {Becker}}, \bibinfo {author} {\bibfnamefont
  {H.}~\bibnamefont {Kaplan}}, \bibinfo {author} {\bibfnamefont {A.~V.}\
  \bibnamefont {Gorshkov}}, \bibinfo {author} {\bibfnamefont {Z.~X.}\
  \bibnamefont {Gong}}, \ and\ \bibinfo {author} {\bibfnamefont
  {C.}~\bibnamefont {Monroe}},\ }\href {\doibase 10.1038/nature24654}
  {\bibfield  {journal} {\bibinfo  {journal} {Nature}\ }\textbf {\bibinfo
  {volume} {551}},\ \bibinfo {pages} {601} (\bibinfo {year}
  {2017})}\BibitemShut {NoStop}%
\bibitem [{\citenamefont {Biamonte}\ \emph {et~al.}(2017)\citenamefont
  {Biamonte}, \citenamefont {Wittek}, \citenamefont {Pancotti}, \citenamefont
  {Rebentrost}, \citenamefont {Wiebe},\ and\ \citenamefont
  {Lloyd}}]{Biamonte2017}%
  \BibitemOpen
  \bibfield  {author} {\bibinfo {author} {\bibfnamefont {J.}~\bibnamefont
  {Biamonte}}, \bibinfo {author} {\bibfnamefont {P.}~\bibnamefont {Wittek}},
  \bibinfo {author} {\bibfnamefont {N.}~\bibnamefont {Pancotti}}, \bibinfo
  {author} {\bibfnamefont {P.}~\bibnamefont {Rebentrost}}, \bibinfo {author}
  {\bibfnamefont {N.}~\bibnamefont {Wiebe}}, \ and\ \bibinfo {author}
  {\bibfnamefont {S.}~\bibnamefont {Lloyd}},\ }\href {\doibase
  10.1038/nature23474} {\bibfield  {journal} {\bibinfo  {journal} {Nature
  Publishing Group}\ }\textbf {\bibinfo {volume} {549}},\ \bibinfo {pages}
  {195} (\bibinfo {year} {2017})}\BibitemShut {NoStop}%
\bibitem [{\citenamefont {Perdomo-Ortiz}\ \emph {et~al.}(2018)\citenamefont
  {Perdomo-Ortiz}, \citenamefont {Benedetti}, \citenamefont
  {Realpe-G{\'{o}}mez},\ and\ \citenamefont {Biswas}}]{Perdomo-Ortiz2018}%
  \BibitemOpen
  \bibfield  {author} {\bibinfo {author} {\bibfnamefont {A.}~\bibnamefont
  {Perdomo-Ortiz}}, \bibinfo {author} {\bibfnamefont {M.}~\bibnamefont
  {Benedetti}}, \bibinfo {author} {\bibfnamefont {J.}~\bibnamefont
  {Realpe-G{\'{o}}mez}}, \ and\ \bibinfo {author} {\bibfnamefont
  {R.}~\bibnamefont {Biswas}},\ }\href@noop {} {\bibfield  {journal} {\bibinfo
  {journal} {Quantum Science and Technology}\ }\textbf {\bibinfo {volume}
  {3}},\ \bibinfo {pages} {030502} (\bibinfo {year} {2018})}\BibitemShut
  {NoStop}%
\bibitem [{\citenamefont {Geirhos}\ \emph {et~al.}(2017)\citenamefont
  {Geirhos}, \citenamefont {Janssen}, \citenamefont {Sch{\"{u}}tt},
  \citenamefont {Rauber}, \citenamefont {Bethge},\ and\ \citenamefont
  {Wichmann}}]{Geirhos2017}%
  \BibitemOpen
  \bibfield  {author} {\bibinfo {author} {\bibfnamefont {R.}~\bibnamefont
  {Geirhos}}, \bibinfo {author} {\bibfnamefont {D.~H.~J.}\ \bibnamefont
  {Janssen}}, \bibinfo {author} {\bibfnamefont {H.~H.}\ \bibnamefont
  {Sch{\"{u}}tt}}, \bibinfo {author} {\bibfnamefont {J.}~\bibnamefont
  {Rauber}}, \bibinfo {author} {\bibfnamefont {M.}~\bibnamefont {Bethge}}, \
  and\ \bibinfo {author} {\bibfnamefont {F.~A.}\ \bibnamefont {Wichmann}},\
  }\href {http://arxiv.org/abs/1706.06969} {\bibfield  {journal} {\bibinfo
  {journal} {arXiv:1706.06969}\ } (\bibinfo {year} {2017})}\BibitemShut
  {NoStop}%
\bibitem [{\citenamefont {Markovic}\ \emph {et~al.}(2020)\citenamefont
  {Markovic}, \citenamefont {Mizrahi}, \citenamefont {Querlioz},\ and\
  \citenamefont {Grollier}}]{Markovic2020a}%
  \BibitemOpen
  \bibfield  {author} {\bibinfo {author} {\bibfnamefont {D.}~\bibnamefont
  {Markovic}}, \bibinfo {author} {\bibfnamefont {A.}~\bibnamefont {Mizrahi}},
  \bibinfo {author} {\bibfnamefont {D.}~\bibnamefont {Querlioz}}, \ and\
  \bibinfo {author} {\bibfnamefont {J.}~\bibnamefont {Grollier}},\ }\href
  {http://arxiv.org/abs/2003.04711} {\bibfield  {journal} {\bibinfo  {journal}
  {arXiv:2003.04711}\ } (\bibinfo {year} {2020})}\BibitemShut {NoStop}%
\bibitem [{\citenamefont {Romera}\ \emph {et~al.}(2018)\citenamefont {Romera},
  \citenamefont {Talatchian}, \citenamefont {Tsunegi}, \citenamefont {Araujo},
  \citenamefont {Cros}, \citenamefont {Bortolotti}, \citenamefont {Yakushiji},
  \citenamefont {Fukushima}, \citenamefont {Kubota}, \citenamefont {Yuasa},
  \citenamefont {Vodenicarevic}, \citenamefont {Locatelli}, \citenamefont
  {Querlioz},\ and\ \citenamefont {Grollier}}]{Romera2017a}%
  \BibitemOpen
  \bibfield  {author} {\bibinfo {author} {\bibfnamefont {M.}~\bibnamefont
  {Romera}}, \bibinfo {author} {\bibfnamefont {P.}~\bibnamefont {Talatchian}},
  \bibinfo {author} {\bibfnamefont {S.}~\bibnamefont {Tsunegi}}, \bibinfo
  {author} {\bibfnamefont {F.~A.}\ \bibnamefont {Araujo}}, \bibinfo {author}
  {\bibfnamefont {V.}~\bibnamefont {Cros}}, \bibinfo {author} {\bibfnamefont
  {P.}~\bibnamefont {Bortolotti}}, \bibinfo {author} {\bibfnamefont
  {K.}~\bibnamefont {Yakushiji}}, \bibinfo {author} {\bibfnamefont
  {A.}~\bibnamefont {Fukushima}}, \bibinfo {author} {\bibfnamefont
  {H.}~\bibnamefont {Kubota}}, \bibinfo {author} {\bibfnamefont
  {S.}~\bibnamefont {Yuasa}}, \bibinfo {author} {\bibfnamefont
  {D.}~\bibnamefont {Vodenicarevic}}, \bibinfo {author} {\bibfnamefont
  {N.}~\bibnamefont {Locatelli}}, \bibinfo {author} {\bibfnamefont
  {D.}~\bibnamefont {Querlioz}}, \ and\ \bibinfo {author} {\bibfnamefont
  {J.}~\bibnamefont {Grollier}},\ }\href {\doibase 10.1038/s41586-018-0632-y}
  {\bibfield  {journal} {\bibinfo  {journal} {Nature}\ }\textbf {\bibinfo
  {volume} {563}},\ \bibinfo {pages} {230} (\bibinfo {year}
  {2018})}\BibitemShut {NoStop}%
\bibitem [{\citenamefont {Feldmann}\ \emph {et~al.}(2019)\citenamefont
  {Feldmann}, \citenamefont {Youngblood}, \citenamefont {Wright},\ and\
  \citenamefont {Bhaskaran}}]{Feldmann2019}%
  \BibitemOpen
  \bibfield  {author} {\bibinfo {author} {\bibfnamefont {J.}~\bibnamefont
  {Feldmann}}, \bibinfo {author} {\bibfnamefont {N.}~\bibnamefont
  {Youngblood}}, \bibinfo {author} {\bibfnamefont {C.~D.}\ \bibnamefont
  {Wright}}, \ and\ \bibinfo {author} {\bibfnamefont {H.}~\bibnamefont
  {Bhaskaran}},\ }\href@noop {} {\bibfield  {journal} {\bibinfo  {journal}
  {Nature}\ }\textbf {\bibinfo {volume} {569}},\ \bibinfo {pages} {208}
  (\bibinfo {year} {2019})}\BibitemShut {NoStop}%
\bibitem [{\citenamefont {Appeltant}\ \emph {et~al.}(2011)\citenamefont
  {Appeltant}, \citenamefont {Soriano}, \citenamefont {Danckaert},
  \citenamefont {Massar}, \citenamefont {Dambre}, \citenamefont {Schrauwen},
  \citenamefont {Mirasso},\ and\ \citenamefont {Fischer}}]{Appeltant2011}%
  \BibitemOpen
  \bibfield  {author} {\bibinfo {author} {\bibfnamefont {L.}~\bibnamefont
  {Appeltant}}, \bibinfo {author} {\bibfnamefont {M.~C.}\ \bibnamefont
  {Soriano}}, \bibinfo {author} {\bibfnamefont {J.}~\bibnamefont {Danckaert}},
  \bibinfo {author} {\bibfnamefont {S.}~\bibnamefont {Massar}}, \bibinfo
  {author} {\bibfnamefont {J.}~\bibnamefont {Dambre}}, \bibinfo {author}
  {\bibfnamefont {B.}~\bibnamefont {Schrauwen}}, \bibinfo {author}
  {\bibfnamefont {C.~R.}\ \bibnamefont {Mirasso}}, \ and\ \bibinfo {author}
  {\bibfnamefont {I.}~\bibnamefont {Fischer}},\ }\href@noop {} {\bibfield
  {journal} {\bibinfo  {journal} {Nature Communications}\ }\textbf {\bibinfo
  {volume} {2}},\ \bibinfo {pages} {468} (\bibinfo {year} {2011})}\BibitemShut
  {NoStop}%
\bibitem [{\citenamefont {Paquot}\ \emph {et~al.}(2012)\citenamefont {Paquot},
  \citenamefont {Duport}, \citenamefont {Smerieri}, \citenamefont {Dambre},
  \citenamefont {Schrauwen}, \citenamefont {Haelterman},\ and\ \citenamefont
  {Massar}}]{Paquot2012}%
  \BibitemOpen
  \bibfield  {author} {\bibinfo {author} {\bibfnamefont {Y.}~\bibnamefont
  {Paquot}}, \bibinfo {author} {\bibfnamefont {F.}~\bibnamefont {Duport}},
  \bibinfo {author} {\bibfnamefont {A.}~\bibnamefont {Smerieri}}, \bibinfo
  {author} {\bibfnamefont {J.}~\bibnamefont {Dambre}}, \bibinfo {author}
  {\bibfnamefont {B.}~\bibnamefont {Schrauwen}}, \bibinfo {author}
  {\bibfnamefont {M.}~\bibnamefont {Haelterman}}, \ and\ \bibinfo {author}
  {\bibfnamefont {S.}~\bibnamefont {Massar}},\ }\href@noop {} {\bibfield
  {journal} {\bibinfo  {journal} {Scientific Reports}\ }\textbf {\bibinfo
  {volume} {2}},\ \bibinfo {pages} {468} (\bibinfo {year} {2012})}\BibitemShut
  {NoStop}%
\bibitem [{\citenamefont {Torrejon}\ \emph {et~al.}(2017)\citenamefont
  {Torrejon}, \citenamefont {Riou}, \citenamefont {Araujo}, \citenamefont
  {Tsunegi}, \citenamefont {Khalsa}, \citenamefont {Querlioz}, \citenamefont
  {Bortolotti}, \citenamefont {Cros}, \citenamefont {Yakushiji}, \citenamefont
  {Fukushima}, \citenamefont {Kubota}, \citenamefont {Yuasa}, \citenamefont
  {Stiles},\ and\ \citenamefont {Grollier}}]{Torrejon2017}%
  \BibitemOpen
  \bibfield  {author} {\bibinfo {author} {\bibfnamefont {J.}~\bibnamefont
  {Torrejon}}, \bibinfo {author} {\bibfnamefont {M.}~\bibnamefont {Riou}},
  \bibinfo {author} {\bibfnamefont {F.~A.}\ \bibnamefont {Araujo}}, \bibinfo
  {author} {\bibfnamefont {S.}~\bibnamefont {Tsunegi}}, \bibinfo {author}
  {\bibfnamefont {G.}~\bibnamefont {Khalsa}}, \bibinfo {author} {\bibfnamefont
  {D.}~\bibnamefont {Querlioz}}, \bibinfo {author} {\bibfnamefont
  {P.}~\bibnamefont {Bortolotti}}, \bibinfo {author} {\bibfnamefont
  {V.}~\bibnamefont {Cros}}, \bibinfo {author} {\bibfnamefont {K.}~\bibnamefont
  {Yakushiji}}, \bibinfo {author} {\bibfnamefont {A.}~\bibnamefont
  {Fukushima}}, \bibinfo {author} {\bibfnamefont {H.}~\bibnamefont {Kubota}},
  \bibinfo {author} {\bibfnamefont {S.}~\bibnamefont {Yuasa}}, \bibinfo
  {author} {\bibfnamefont {M.~D.}\ \bibnamefont {Stiles}}, \ and\ \bibinfo
  {author} {\bibfnamefont {J.}~\bibnamefont {Grollier}},\ }\href@noop {}
  {\bibfield  {journal} {\bibinfo  {journal} {Nature}\ }\textbf {\bibinfo
  {volume} {547}},\ \bibinfo {pages} {428} (\bibinfo {year}
  {2017})}\BibitemShut {NoStop}%
\bibitem [{\citenamefont {Markovi\'c}\ \emph {et~al.}(2019)\citenamefont
  {Markovi\'c}, \citenamefont {Leroux}, \citenamefont {Riou}, \citenamefont
  {{Abreu Araujo}}, \citenamefont {Torrejon}, \citenamefont {Querlioz},
  \citenamefont {Fukushima}, \citenamefont {Yasa}, \citenamefont {Trastoy},
  \citenamefont {Bortolotti},\ and\ \citenamefont {Grollier}}]{Markovic2019}%
  \BibitemOpen
  \bibfield  {author} {\bibinfo {author} {\bibfnamefont {D.}~\bibnamefont
  {Markovi\'c}}, \bibinfo {author} {\bibfnamefont {N.}~\bibnamefont {Leroux}},
  \bibinfo {author} {\bibfnamefont {M.}~\bibnamefont {Riou}}, \bibinfo {author}
  {\bibfnamefont {F.}~\bibnamefont {{Abreu Araujo}}}, \bibinfo {author}
  {\bibfnamefont {J.}~\bibnamefont {Torrejon}}, \bibinfo {author}
  {\bibfnamefont {D.}~\bibnamefont {Querlioz}}, \bibinfo {author}
  {\bibfnamefont {A.}~\bibnamefont {Fukushima}}, \bibinfo {author}
  {\bibfnamefont {S.}~\bibnamefont {Yasa}}, \bibinfo {author} {\bibfnamefont
  {J.}~\bibnamefont {Trastoy}}, \bibinfo {author} {\bibfnamefont
  {P.}~\bibnamefont {Bortolotti}}, \ and\ \bibinfo {author} {\bibfnamefont
  {J.}~\bibnamefont {Grollier}},\ }\href {\doibase 10.1063/1.5079305}
  {\bibfield  {journal} {\bibinfo  {journal} {Applied Physics Letters}\
  }\textbf {\bibinfo {volume} {114}},\ \bibinfo {pages} {012409} (\bibinfo
  {year} {2019})}\BibitemShut {NoStop}%
\bibitem [{\citenamefont {Benedetti}\ \emph
  {et~al.}(2019{\natexlab{a}})\citenamefont {Benedetti}, \citenamefont {Lloyd},
  \citenamefont {Sack},\ and\ \citenamefont {Fiorentini}}]{Benedetti2019a}%
  \BibitemOpen
  \bibfield  {author} {\bibinfo {author} {\bibfnamefont {M.}~\bibnamefont
  {Benedetti}}, \bibinfo {author} {\bibfnamefont {E.}~\bibnamefont {Lloyd}},
  \bibinfo {author} {\bibfnamefont {S.}~\bibnamefont {Sack}}, \ and\ \bibinfo
  {author} {\bibfnamefont {M.}~\bibnamefont {Fiorentini}},\ }\href {\doibase
  10.1088/2058-9565/ab4eb5} {\bibfield  {journal} {\bibinfo  {journal} {Quantum
  Science and Technology}\ }\textbf {\bibinfo {volume} {4}},\ \bibinfo {pages}
  {043001} (\bibinfo {year} {2019}{\natexlab{a}})}\BibitemShut {NoStop}%
\bibitem [{\citenamefont {Tacchino}\ \emph {et~al.}(2019)\citenamefont
  {Tacchino}, \citenamefont {Macchiavello}, \citenamefont {Gerace},\ and\
  \citenamefont {Bajoni}}]{Tacchino2019}%
  \BibitemOpen
  \bibfield  {author} {\bibinfo {author} {\bibfnamefont {F.}~\bibnamefont
  {Tacchino}}, \bibinfo {author} {\bibfnamefont {C.}~\bibnamefont
  {Macchiavello}}, \bibinfo {author} {\bibfnamefont {D.}~\bibnamefont
  {Gerace}}, \ and\ \bibinfo {author} {\bibfnamefont {D.}~\bibnamefont
  {Bajoni}},\ }\href {\doibase 10.1038/s41534-019-0140-4} {\bibfield  {journal}
  {\bibinfo  {journal} {npj Quantum Information}\ }\textbf {\bibinfo {volume}
  {5}},\ \bibinfo {pages} {1} (\bibinfo {year} {2019})}\BibitemShut {NoStop}%
\bibitem [{\citenamefont {Stoudenmire}\ and\ \citenamefont
  {Schwab}(2016)}]{Stoudenmire2016}%
  \BibitemOpen
  \bibfield  {author} {\bibinfo {author} {\bibfnamefont {E.~M.}\ \bibnamefont
  {Stoudenmire}}\ and\ \bibinfo {author} {\bibfnamefont {D.~J.}\ \bibnamefont
  {Schwab}},\ }\href@noop {} {\bibfield  {journal} {\bibinfo  {journal}
  {Advances in Neural Information Processing Systems}\ }\textbf {\bibinfo
  {volume} {29}},\ \bibinfo {pages} {4806} (\bibinfo {year}
  {2016})}\BibitemShut {NoStop}%
\bibitem [{\citenamefont {Mitarai}\ \emph {et~al.}(2018)\citenamefont
  {Mitarai}, \citenamefont {Negoro}, \citenamefont {Kitagawa},\ and\
  \citenamefont {Fujii}}]{Mitarai2018}%
  \BibitemOpen
  \bibfield  {author} {\bibinfo {author} {\bibfnamefont {K.}~\bibnamefont
  {Mitarai}}, \bibinfo {author} {\bibfnamefont {M.}~\bibnamefont {Negoro}},
  \bibinfo {author} {\bibfnamefont {M.}~\bibnamefont {Kitagawa}}, \ and\
  \bibinfo {author} {\bibfnamefont {K.}~\bibnamefont {Fujii}},\ }\href
  {\doibase 10.1103/PhysRevA.98.032309} {\bibfield  {journal} {\bibinfo
  {journal} {Physical Review A}\ }\textbf {\bibinfo {volume} {98}},\ \bibinfo
  {pages} {1} (\bibinfo {year} {2018})}\BibitemShut {NoStop}%
\bibitem [{\citenamefont {Grant}\ \emph {et~al.}(2018)\citenamefont {Grant},
  \citenamefont {Benedetti}, \citenamefont {Cao}, \citenamefont {Hallam},
  \citenamefont {Lockhart}, \citenamefont {Stojevic}, \citenamefont {Green},\
  and\ \citenamefont {Severini}}]{Grant2018}%
  \BibitemOpen
  \bibfield  {author} {\bibinfo {author} {\bibfnamefont {E.}~\bibnamefont
  {Grant}}, \bibinfo {author} {\bibfnamefont {M.}~\bibnamefont {Benedetti}},
  \bibinfo {author} {\bibfnamefont {S.}~\bibnamefont {Cao}}, \bibinfo {author}
  {\bibfnamefont {A.}~\bibnamefont {Hallam}}, \bibinfo {author} {\bibfnamefont
  {J.}~\bibnamefont {Lockhart}}, \bibinfo {author} {\bibfnamefont
  {V.}~\bibnamefont {Stojevic}}, \bibinfo {author} {\bibfnamefont {A.~G.}\
  \bibnamefont {Green}}, \ and\ \bibinfo {author} {\bibfnamefont
  {S.}~\bibnamefont {Severini}},\ }\href {\doibase 10.1038/s41534-018-0116-9}
  {\bibfield  {journal} {\bibinfo  {journal} {npj Quantum Information}\
  }\textbf {\bibinfo {volume} {4}},\ \bibinfo {pages} {17} (\bibinfo {year}
  {2018})}\BibitemShut {NoStop}%
\bibitem [{\citenamefont {Havl{\'{i}}{\v{c}}ek}\ \emph
  {et~al.}(2019)\citenamefont {Havl{\'{i}}{\v{c}}ek}, \citenamefont
  {C{\'{o}}rcoles}, \citenamefont {Temme}, \citenamefont {Harrow},
  \citenamefont {Kandala}, \citenamefont {Chow},\ and\ \citenamefont
  {Gambetta}}]{Havlicek2019}%
  \BibitemOpen
  \bibfield  {author} {\bibinfo {author} {\bibfnamefont {V.}~\bibnamefont
  {Havl{\'{i}}{\v{c}}ek}}, \bibinfo {author} {\bibfnamefont {A.~D.}\
  \bibnamefont {C{\'{o}}rcoles}}, \bibinfo {author} {\bibfnamefont
  {K.}~\bibnamefont {Temme}}, \bibinfo {author} {\bibfnamefont {A.~W.}\
  \bibnamefont {Harrow}}, \bibinfo {author} {\bibfnamefont {A.}~\bibnamefont
  {Kandala}}, \bibinfo {author} {\bibfnamefont {J.~M.}\ \bibnamefont {Chow}}, \
  and\ \bibinfo {author} {\bibfnamefont {J.~M.}\ \bibnamefont {Gambetta}},\
  }\href {\doibase 10.1038/s41586-019-0980-2} {\bibfield  {journal} {\bibinfo
  {journal} {Nature}\ }\textbf {\bibinfo {volume} {567}},\ \bibinfo {pages}
  {209} (\bibinfo {year} {2019})}\BibitemShut {NoStop}%
\bibitem [{\citenamefont {Henderson}\ \emph {et~al.}(2020)\citenamefont
  {Henderson}, \citenamefont {Shakya}, \citenamefont {Pradhan},\ and\
  \citenamefont {Cook}}]{Henderson2020}%
  \BibitemOpen
  \bibfield  {author} {\bibinfo {author} {\bibfnamefont {M.}~\bibnamefont
  {Henderson}}, \bibinfo {author} {\bibfnamefont {S.}~\bibnamefont {Shakya}},
  \bibinfo {author} {\bibfnamefont {S.}~\bibnamefont {Pradhan}}, \ and\
  \bibinfo {author} {\bibfnamefont {T.}~\bibnamefont {Cook}},\ }\href@noop {}
  {\bibfield  {journal} {\bibinfo  {journal} {Quantum Machine Intelligence}\
  }\textbf {\bibinfo {volume} {2}},\ \bibinfo {pages} {1} (\bibinfo {year}
  {2020})}\BibitemShut {NoStop}%
\bibitem [{\citenamefont {Han}\ \emph {et~al.}(2018)\citenamefont {Han},
  \citenamefont {Wang}, \citenamefont {Fan}, \citenamefont {Wang},\ and\
  \citenamefont {Zhang}}]{Han2018}%
  \BibitemOpen
  \bibfield  {author} {\bibinfo {author} {\bibfnamefont {Z.~Y.}\ \bibnamefont
  {Han}}, \bibinfo {author} {\bibfnamefont {J.}~\bibnamefont {Wang}}, \bibinfo
  {author} {\bibfnamefont {H.}~\bibnamefont {Fan}}, \bibinfo {author}
  {\bibfnamefont {L.}~\bibnamefont {Wang}}, \ and\ \bibinfo {author}
  {\bibfnamefont {P.}~\bibnamefont {Zhang}},\ }\href
  {https://doi.org/10.1103/PhysRevX.8.031012} {\bibfield  {journal} {\bibinfo
  {journal} {Physical Review X}\ }\textbf {\bibinfo {volume} {8}},\ \bibinfo
  {pages} {31012} (\bibinfo {year} {2018})}\BibitemShut {NoStop}%
\bibitem [{\citenamefont {Mohamed}\ and\ \citenamefont
  {Lakshminarayanan}(2016)}]{Mohamed2016}%
  \BibitemOpen
  \bibfield  {author} {\bibinfo {author} {\bibfnamefont {S.}~\bibnamefont
  {Mohamed}}\ and\ \bibinfo {author} {\bibfnamefont {B.}~\bibnamefont
  {Lakshminarayanan}},\ }\href {http://arxiv.org/abs/1610.03483} {\bibfield
  {journal} {\bibinfo  {journal} {arXiv 1610.03483}\ } (\bibinfo {year}
  {2016})}\BibitemShut {NoStop}%
\bibitem [{\citenamefont {Hamilton}, \citenamefont {Dumitrescu},\ and\
  \citenamefont {Pooser}(2019)}]{Hamilton2019}%
  \BibitemOpen
  \bibfield  {author} {\bibinfo {author} {\bibfnamefont {K.~E.}\ \bibnamefont
  {Hamilton}}, \bibinfo {author} {\bibfnamefont {E.~F.}\ \bibnamefont
  {Dumitrescu}}, \ and\ \bibinfo {author} {\bibfnamefont {R.~C.}\ \bibnamefont
  {Pooser}},\ }\href {\doibase 10.1103/PhysRevA.99.062323} {\bibfield
  {journal} {\bibinfo  {journal} {Physical Review A}\ }\textbf {\bibinfo
  {volume} {99}},\ \bibinfo {pages} {1} (\bibinfo {year} {2019})}\BibitemShut
  {NoStop}%
\bibitem [{\citenamefont {Benedetti}\ \emph
  {et~al.}(2019{\natexlab{b}})\citenamefont {Benedetti}, \citenamefont
  {Garcia-Pintos}, \citenamefont {Perdomo}, \citenamefont {Leyton-Ortega},
  \citenamefont {Nam},\ and\ \citenamefont {Perdomo-Ortiz}}]{Benedetti2019b}%
  \BibitemOpen
  \bibfield  {author} {\bibinfo {author} {\bibfnamefont {M.}~\bibnamefont
  {Benedetti}}, \bibinfo {author} {\bibfnamefont {D.}~\bibnamefont
  {Garcia-Pintos}}, \bibinfo {author} {\bibfnamefont {O.}~\bibnamefont
  {Perdomo}}, \bibinfo {author} {\bibfnamefont {V.}~\bibnamefont
  {Leyton-Ortega}}, \bibinfo {author} {\bibfnamefont {Y.}~\bibnamefont {Nam}},
  \ and\ \bibinfo {author} {\bibfnamefont {A.}~\bibnamefont {Perdomo-Ortiz}},\
  }\href@noop {} {\bibfield  {journal} {\bibinfo  {journal} {npj Quantum
  Information}\ }\textbf {\bibinfo {volume} {5}} (\bibinfo {year}
  {2019}{\natexlab{b}})}\BibitemShut {NoStop}%
\bibitem [{\citenamefont {Leyton-Ortega}, \citenamefont {Perdomo-Ortiz},\ and\
  \citenamefont {Perdomo}(2019)}]{Leyton-Ortega2019}%
  \BibitemOpen
  \bibfield  {author} {\bibinfo {author} {\bibfnamefont {V.}~\bibnamefont
  {Leyton-Ortega}}, \bibinfo {author} {\bibfnamefont {A.}~\bibnamefont
  {Perdomo-Ortiz}}, \ and\ \bibinfo {author} {\bibfnamefont {O.}~\bibnamefont
  {Perdomo}},\ }\href {http://arxiv.org/abs/1901.08047} {\bibfield  {journal}
  {\bibinfo  {journal} {arXiv:1901.08047v1}\ } (\bibinfo {year}
  {2019})}\BibitemShut {NoStop}%
\bibitem [{\citenamefont {Rocchetto}\ \emph {et~al.}(2019)\citenamefont
  {Rocchetto}, \citenamefont {Aaronson}, \citenamefont {Severini},
  \citenamefont {Carvacho}, \citenamefont {Poderini}, \citenamefont {Agresti},
  \citenamefont {Bentivegna},\ and\ \citenamefont {Sciarrino}}]{Rocchetto2019}%
  \BibitemOpen
  \bibfield  {author} {\bibinfo {author} {\bibfnamefont {A.}~\bibnamefont
  {Rocchetto}}, \bibinfo {author} {\bibfnamefont {S.}~\bibnamefont {Aaronson}},
  \bibinfo {author} {\bibfnamefont {S.}~\bibnamefont {Severini}}, \bibinfo
  {author} {\bibfnamefont {G.}~\bibnamefont {Carvacho}}, \bibinfo {author}
  {\bibfnamefont {D.}~\bibnamefont {Poderini}}, \bibinfo {author}
  {\bibfnamefont {I.}~\bibnamefont {Agresti}}, \bibinfo {author} {\bibfnamefont
  {M.}~\bibnamefont {Bentivegna}}, \ and\ \bibinfo {author} {\bibfnamefont
  {F.}~\bibnamefont {Sciarrino}},\ }\href {\doibase 10.1126/sciadv.aau1946}
  {\bibfield  {journal} {\bibinfo  {journal} {Science Advances}\ }\textbf
  {\bibinfo {volume} {5}},\ \bibinfo {pages} {1} (\bibinfo {year}
  {2019})}\BibitemShut {NoStop}%
\bibitem [{\citenamefont {Cong}, \citenamefont {Choi},\ and\ \citenamefont
  {Lukin}(2019)}]{Cong2019}%
  \BibitemOpen
  \bibfield  {author} {\bibinfo {author} {\bibfnamefont {I.}~\bibnamefont
  {Cong}}, \bibinfo {author} {\bibfnamefont {S.}~\bibnamefont {Choi}}, \ and\
  \bibinfo {author} {\bibfnamefont {M.~D.}\ \bibnamefont {Lukin}},\ }\href
  {http://dx.doi.org/10.1038/s41567-019-0648-8} {\bibfield  {journal} {\bibinfo
   {journal} {Nature Physics}\ }\textbf {\bibinfo {volume} {15}},\ \bibinfo
  {pages} {1273} (\bibinfo {year} {2019})}\BibitemShut {NoStop}%
\bibitem [{\citenamefont {Chen}\ \emph {et~al.}(2018)\citenamefont {Chen},
  \citenamefont {Wossnig}, \citenamefont {Severini}, \citenamefont {Neven},\
  and\ \citenamefont {Mohseni}}]{Chen2018a}%
  \BibitemOpen
  \bibfield  {author} {\bibinfo {author} {\bibfnamefont {H.}~\bibnamefont
  {Chen}}, \bibinfo {author} {\bibfnamefont {L.}~\bibnamefont {Wossnig}},
  \bibinfo {author} {\bibfnamefont {S.}~\bibnamefont {Severini}}, \bibinfo
  {author} {\bibfnamefont {H.}~\bibnamefont {Neven}}, \ and\ \bibinfo {author}
  {\bibfnamefont {M.}~\bibnamefont {Mohseni}},\ }\href
  {http://arxiv.org/abs/1805.08654} {\bibfield  {journal} {\bibinfo  {journal}
  {arXiv:1805.08654}\ } (\bibinfo {year} {2018})}\BibitemShut {NoStop}%
\bibitem [{\citenamefont {Lloyd}\ and\ \citenamefont
  {Weedbrook}(2018)}]{Lloyd2018}%
  \BibitemOpen
  \bibfield  {author} {\bibinfo {author} {\bibfnamefont {S.}~\bibnamefont
  {Lloyd}}\ and\ \bibinfo {author} {\bibfnamefont {C.}~\bibnamefont
  {Weedbrook}},\ }\href {\doibase 10.1103/PhysRevLett.121.040502} {\bibfield
  {journal} {\bibinfo  {journal} {Physical Review Letters}\ }\textbf {\bibinfo
  {volume} {121}},\ \bibinfo {pages} {40502} (\bibinfo {year}
  {2018})}\BibitemShut {NoStop}%
\bibitem [{\citenamefont {Hu}\ \emph {et~al.}(2019)\citenamefont {Hu},
  \citenamefont {Wu}, \citenamefont {Cai}, \citenamefont {Ma}, \citenamefont
  {Mu}, \citenamefont {Xu}, \citenamefont {Wang}, \citenamefont {Song},
  \citenamefont {Deng}, \citenamefont {Zou},\ and\ \citenamefont
  {Sun}}]{Hu2019}%
  \BibitemOpen
  \bibfield  {author} {\bibinfo {author} {\bibfnamefont {L.}~\bibnamefont
  {Hu}}, \bibinfo {author} {\bibfnamefont {S.~H.}\ \bibnamefont {Wu}}, \bibinfo
  {author} {\bibfnamefont {W.}~\bibnamefont {Cai}}, \bibinfo {author}
  {\bibfnamefont {Y.}~\bibnamefont {Ma}}, \bibinfo {author} {\bibfnamefont
  {X.}~\bibnamefont {Mu}}, \bibinfo {author} {\bibfnamefont {Y.}~\bibnamefont
  {Xu}}, \bibinfo {author} {\bibfnamefont {H.}~\bibnamefont {Wang}}, \bibinfo
  {author} {\bibfnamefont {Y.}~\bibnamefont {Song}}, \bibinfo {author}
  {\bibfnamefont {D.~L.}\ \bibnamefont {Deng}}, \bibinfo {author}
  {\bibfnamefont {C.~L.}\ \bibnamefont {Zou}}, \ and\ \bibinfo {author}
  {\bibfnamefont {L.}~\bibnamefont {Sun}},\ }\href {\doibase
  10.1126/sciadv.aav2761} {\bibfield  {journal} {\bibinfo  {journal} {Science
  Advances}\ }\textbf {\bibinfo {volume} {5}},\ \bibinfo {pages} {1} (\bibinfo
  {year} {2019})}\BibitemShut {NoStop}%
\bibitem [{\citenamefont {O'Malley}\ \emph {et~al.}(2016)\citenamefont
  {O'Malley}, \citenamefont {Babbush}, \citenamefont {Kivlichan}, \citenamefont
  {Romero}, \citenamefont {McClean}, \citenamefont {Barends}, \citenamefont
  {Kelly}, \citenamefont {Roushan}, \citenamefont {Tranter}, \citenamefont
  {Ding}, \citenamefont {Campbell}, \citenamefont {Chen}, \citenamefont {Chen},
  \citenamefont {Chiaro}, \citenamefont {Dunsworth}, \citenamefont {Fowler},
  \citenamefont {Jeffrey}, \citenamefont {Lucero}, \citenamefont {Megrant},
  \citenamefont {Mutus}, \citenamefont {Neeley}, \citenamefont {Neill},
  \citenamefont {Quintana}, \citenamefont {Sank}, \citenamefont {Vainsencher},
  \citenamefont {Wenner}, \citenamefont {White}, \citenamefont {Coveney},
  \citenamefont {Love}, \citenamefont {Neven}, \citenamefont {Aspuru-Guzik},\
  and\ \citenamefont {Martinis}}]{OMalley2016}%
  \BibitemOpen
  \bibfield  {author} {\bibinfo {author} {\bibfnamefont {P.~J.}\ \bibnamefont
  {O'Malley}}, \bibinfo {author} {\bibfnamefont {R.}~\bibnamefont {Babbush}},
  \bibinfo {author} {\bibfnamefont {I.~D.}\ \bibnamefont {Kivlichan}}, \bibinfo
  {author} {\bibfnamefont {J.}~\bibnamefont {Romero}}, \bibinfo {author}
  {\bibfnamefont {J.~R.}\ \bibnamefont {McClean}}, \bibinfo {author}
  {\bibfnamefont {R.}~\bibnamefont {Barends}}, \bibinfo {author} {\bibfnamefont
  {J.}~\bibnamefont {Kelly}}, \bibinfo {author} {\bibfnamefont
  {P.}~\bibnamefont {Roushan}}, \bibinfo {author} {\bibfnamefont
  {A.}~\bibnamefont {Tranter}}, \bibinfo {author} {\bibfnamefont
  {N.}~\bibnamefont {Ding}}, \bibinfo {author} {\bibfnamefont {B.}~\bibnamefont
  {Campbell}}, \bibinfo {author} {\bibfnamefont {Y.}~\bibnamefont {Chen}},
  \bibinfo {author} {\bibfnamefont {Z.}~\bibnamefont {Chen}}, \bibinfo {author}
  {\bibfnamefont {B.}~\bibnamefont {Chiaro}}, \bibinfo {author} {\bibfnamefont
  {A.}~\bibnamefont {Dunsworth}}, \bibinfo {author} {\bibfnamefont {A.~G.}\
  \bibnamefont {Fowler}}, \bibinfo {author} {\bibfnamefont {E.}~\bibnamefont
  {Jeffrey}}, \bibinfo {author} {\bibfnamefont {E.}~\bibnamefont {Lucero}},
  \bibinfo {author} {\bibfnamefont {A.}~\bibnamefont {Megrant}}, \bibinfo
  {author} {\bibfnamefont {J.~Y.}\ \bibnamefont {Mutus}}, \bibinfo {author}
  {\bibfnamefont {M.}~\bibnamefont {Neeley}}, \bibinfo {author} {\bibfnamefont
  {C.}~\bibnamefont {Neill}}, \bibinfo {author} {\bibfnamefont
  {C.}~\bibnamefont {Quintana}}, \bibinfo {author} {\bibfnamefont
  {D.}~\bibnamefont {Sank}}, \bibinfo {author} {\bibfnamefont {A.}~\bibnamefont
  {Vainsencher}}, \bibinfo {author} {\bibfnamefont {J.}~\bibnamefont {Wenner}},
  \bibinfo {author} {\bibfnamefont {T.~C.}\ \bibnamefont {White}}, \bibinfo
  {author} {\bibfnamefont {P.~V.}\ \bibnamefont {Coveney}}, \bibinfo {author}
  {\bibfnamefont {P.~J.}\ \bibnamefont {Love}}, \bibinfo {author}
  {\bibfnamefont {H.}~\bibnamefont {Neven}}, \bibinfo {author} {\bibfnamefont
  {A.}~\bibnamefont {Aspuru-Guzik}}, \ and\ \bibinfo {author} {\bibfnamefont
  {J.~M.}\ \bibnamefont {Martinis}},\ }\href@noop {} {\bibfield  {journal}
  {\bibinfo  {journal} {Physical Review X}\ }\textbf {\bibinfo {volume} {6}},\
  \bibinfo {pages} {1} (\bibinfo {year} {2016})}\BibitemShut {NoStop}%
\bibitem [{\citenamefont {Hempel}\ \emph {et~al.}(2018)\citenamefont {Hempel},
  \citenamefont {Maier}, \citenamefont {Romero}, \citenamefont {McClean},
  \citenamefont {Monz}, \citenamefont {Shen}, \citenamefont {Jurcevic},
  \citenamefont {Lanyon}, \citenamefont {Love}, \citenamefont {Babbush},
  \citenamefont {Aspuru-Guzik}, \citenamefont {Blatt},\ and\ \citenamefont
  {Roos}}]{Hempel2018}%
  \BibitemOpen
  \bibfield  {author} {\bibinfo {author} {\bibfnamefont {C.}~\bibnamefont
  {Hempel}}, \bibinfo {author} {\bibfnamefont {C.}~\bibnamefont {Maier}},
  \bibinfo {author} {\bibfnamefont {J.}~\bibnamefont {Romero}}, \bibinfo
  {author} {\bibfnamefont {J.}~\bibnamefont {McClean}}, \bibinfo {author}
  {\bibfnamefont {T.}~\bibnamefont {Monz}}, \bibinfo {author} {\bibfnamefont
  {H.}~\bibnamefont {Shen}}, \bibinfo {author} {\bibfnamefont {P.}~\bibnamefont
  {Jurcevic}}, \bibinfo {author} {\bibfnamefont {B.~P.}\ \bibnamefont
  {Lanyon}}, \bibinfo {author} {\bibfnamefont {P.}~\bibnamefont {Love}},
  \bibinfo {author} {\bibfnamefont {R.}~\bibnamefont {Babbush}}, \bibinfo
  {author} {\bibfnamefont {A.}~\bibnamefont {Aspuru-Guzik}}, \bibinfo {author}
  {\bibfnamefont {R.}~\bibnamefont {Blatt}}, \ and\ \bibinfo {author}
  {\bibfnamefont {C.~F.}\ \bibnamefont {Roos}},\ }\href
  {https://doi.org/10.1103/PhysRevX.8.031022} {\bibfield  {journal} {\bibinfo
  {journal} {Physical Review X}\ }\textbf {\bibinfo {volume} {8}},\ \bibinfo
  {pages} {31022} (\bibinfo {year} {2018})}\BibitemShut {NoStop}%
\bibitem [{\citenamefont {Morales}, \citenamefont {Tlyachev},\ and\
  \citenamefont {Biamonte}(2018)}]{Morales2018}%
  \BibitemOpen
  \bibfield  {author} {\bibinfo {author} {\bibfnamefont {M.~E.}\ \bibnamefont
  {Morales}}, \bibinfo {author} {\bibfnamefont {T.}~\bibnamefont {Tlyachev}}, \
  and\ \bibinfo {author} {\bibfnamefont {J.}~\bibnamefont {Biamonte}},\ }\href
  {\doibase 10.1103/PhysRevA.98.062333} {\bibfield  {journal} {\bibinfo
  {journal} {Physical Review A}\ }\textbf {\bibinfo {volume} {98}},\ \bibinfo
  {pages} {1} (\bibinfo {year} {2018})}\BibitemShut {NoStop}%
\bibitem [{\citenamefont {Benedetti}\ \emph {et~al.}(2017)\citenamefont
  {Benedetti}, \citenamefont {Realpe-G{\'{o}}mez}, \citenamefont {Biswas},\
  and\ \citenamefont {Perdomo-Ortiz}}]{Benedetti2017}%
  \BibitemOpen
  \bibfield  {author} {\bibinfo {author} {\bibfnamefont {M.}~\bibnamefont
  {Benedetti}}, \bibinfo {author} {\bibfnamefont {J.}~\bibnamefont
  {Realpe-G{\'{o}}mez}}, \bibinfo {author} {\bibfnamefont {R.}~\bibnamefont
  {Biswas}}, \ and\ \bibinfo {author} {\bibfnamefont {A.}~\bibnamefont
  {Perdomo-Ortiz}},\ }\href {\doibase 10.1103/PhysRevX.7.041052} {\bibfield
  {journal} {\bibinfo  {journal} {Physical Review X}\ }\textbf {\bibinfo
  {volume} {7}},\ \bibinfo {pages} {52} (\bibinfo {year} {2017})}\BibitemShut
  {NoStop}%
\bibitem [{\citenamefont {Amin}\ \emph {et~al.}(2018)\citenamefont {Amin},
  \citenamefont {Andriyash}, \citenamefont {Rolfe}, \citenamefont
  {Kulchytskyy},\ and\ \citenamefont {Melko}}]{Amin2018}%
  \BibitemOpen
  \bibfield  {author} {\bibinfo {author} {\bibfnamefont {M.~H.}\ \bibnamefont
  {Amin}}, \bibinfo {author} {\bibfnamefont {E.}~\bibnamefont {Andriyash}},
  \bibinfo {author} {\bibfnamefont {J.}~\bibnamefont {Rolfe}}, \bibinfo
  {author} {\bibfnamefont {B.}~\bibnamefont {Kulchytskyy}}, \ and\ \bibinfo
  {author} {\bibfnamefont {R.}~\bibnamefont {Melko}},\ }\href
  {https://doi.org/10.1103/PhysRevX.8.021050} {\bibfield  {journal} {\bibinfo
  {journal} {Physical Review X}\ }\textbf {\bibinfo {volume} {8}},\ \bibinfo
  {pages} {21050} (\bibinfo {year} {2018})}\BibitemShut {NoStop}%
\bibitem [{\citenamefont {Fujii}\ and\ \citenamefont
  {Nakajima}(2017)}]{Fujii2017}%
  \BibitemOpen
  \bibfield  {author} {\bibinfo {author} {\bibfnamefont {K.}~\bibnamefont
  {Fujii}}\ and\ \bibinfo {author} {\bibfnamefont {K.}~\bibnamefont
  {Nakajima}},\ }\href@noop {} {\bibfield  {journal} {\bibinfo  {journal}
  {Physical Review Applied}\ }\textbf {\bibinfo {volume} {8}},\ \bibinfo
  {pages} {024030} (\bibinfo {year} {2017})}\BibitemShut {NoStop}%
\bibitem [{\citenamefont {Nakajima}\ \emph {et~al.}(2019)\citenamefont
  {Nakajima}, \citenamefont {Fujii}, \citenamefont {Negoro}, \citenamefont
  {Mitarai},\ and\ \citenamefont {Kitagawa}}]{Nakajima2019}%
  \BibitemOpen
  \bibfield  {author} {\bibinfo {author} {\bibfnamefont {K.}~\bibnamefont
  {Nakajima}}, \bibinfo {author} {\bibfnamefont {K.}~\bibnamefont {Fujii}},
  \bibinfo {author} {\bibfnamefont {M.}~\bibnamefont {Negoro}}, \bibinfo
  {author} {\bibfnamefont {K.}~\bibnamefont {Mitarai}}, \ and\ \bibinfo
  {author} {\bibfnamefont {M.}~\bibnamefont {Kitagawa}},\ }\href {\doibase
  10.1103/PhysRevApplied.11.034021} {\bibfield  {journal} {\bibinfo  {journal}
  {Physical Review Applied}\ }\textbf {\bibinfo {volume} {11}},\ \bibinfo
  {pages} {1} (\bibinfo {year} {2019})}\BibitemShut {NoStop}%
\bibitem [{\citenamefont {Ghosh}\ \emph {et~al.}(2019)\citenamefont {Ghosh},
  \citenamefont {Opala}, \citenamefont {Matuszewski}, \citenamefont {Paterek},\
  and\ \citenamefont {Liew}}]{Ghosh2019}%
  \BibitemOpen
  \bibfield  {author} {\bibinfo {author} {\bibfnamefont {S.}~\bibnamefont
  {Ghosh}}, \bibinfo {author} {\bibfnamefont {A.}~\bibnamefont {Opala}},
  \bibinfo {author} {\bibfnamefont {M.}~\bibnamefont {Matuszewski}}, \bibinfo
  {author} {\bibfnamefont {T.}~\bibnamefont {Paterek}}, \ and\ \bibinfo
  {author} {\bibfnamefont {T.~C.~H.}\ \bibnamefont {Liew}},\ }\href@noop {}
  {\bibfield  {journal} {\bibinfo  {journal} {npj Quantum Information}\
  }\textbf {\bibinfo {volume} {5}},\ \bibinfo {pages} {35} (\bibinfo {year}
  {2019})}\BibitemShut {NoStop}%
\bibitem [{\citenamefont {Chen}, \citenamefont {Nurdin},\ and\ \citenamefont
  {Yamamoto}(2020)}]{Chen2020}%
  \BibitemOpen
  \bibfield  {author} {\bibinfo {author} {\bibfnamefont {J.}~\bibnamefont
  {Chen}}, \bibinfo {author} {\bibfnamefont {H.~I.}\ \bibnamefont {Nurdin}}, \
  and\ \bibinfo {author} {\bibfnamefont {N.}~\bibnamefont {Yamamoto}},\ }\href
  {http://arxiv.org/abs/2001.09498} {\bibfield  {journal} {\bibinfo  {journal}
  {arXiv:2001.09498}\ } (\bibinfo {year} {2020})}\BibitemShut {NoStop}%
\bibitem [{\citenamefont {Dasgupta}\ \emph {et~al.}(2020)\citenamefont
  {Dasgupta}, \citenamefont {Hamilton}, \citenamefont {Lougovski},\ and\
  \citenamefont {Banerjee}}]{Dasgupta2020}%
  \BibitemOpen
  \bibfield  {author} {\bibinfo {author} {\bibfnamefont {S.}~\bibnamefont
  {Dasgupta}}, \bibinfo {author} {\bibfnamefont {K.~E.}\ \bibnamefont
  {Hamilton}}, \bibinfo {author} {\bibfnamefont {P.}~\bibnamefont {Lougovski}},
  \ and\ \bibinfo {author} {\bibfnamefont {A.}~\bibnamefont {Banerjee}},\
  }\href {http://arxiv.org/abs/2004.08240} {\bibfield  {journal} {\bibinfo
  {journal} {arXiv:2004.08240}\ } (\bibinfo {year} {2020})}\BibitemShut
  {NoStop}%
\bibitem [{\citenamefont {Ghosh}, \citenamefont {Paterek},\ and\ \citenamefont
  {Liew}(2019)}]{Ghosh2019b}%
  \BibitemOpen
  \bibfield  {author} {\bibinfo {author} {\bibfnamefont {S.}~\bibnamefont
  {Ghosh}}, \bibinfo {author} {\bibfnamefont {T.}~\bibnamefont {Paterek}}, \
  and\ \bibinfo {author} {\bibfnamefont {T.~C.}\ \bibnamefont {Liew}},\ }\href
  {\doibase 10.1103/PhysRevLett.123.260404} {\bibfield  {journal} {\bibinfo
  {journal} {Physical Review Letters}\ }\textbf {\bibinfo {volume} {123}},\
  \bibinfo {pages} {260404} (\bibinfo {year} {2019})}\BibitemShut {NoStop}%
\bibitem [{\citenamefont {Ghosh}\ \emph {et~al.}(2020)\citenamefont {Ghosh},
  \citenamefont {Krisnanda}, \citenamefont {Paterek},\ and\ \citenamefont
  {Liew}}]{Ghosh2020}%
  \BibitemOpen
  \bibfield  {author} {\bibinfo {author} {\bibfnamefont {S.}~\bibnamefont
  {Ghosh}}, \bibinfo {author} {\bibfnamefont {T.}~\bibnamefont {Krisnanda}},
  \bibinfo {author} {\bibfnamefont {T.}~\bibnamefont {Paterek}}, \ and\
  \bibinfo {author} {\bibfnamefont {T.~C.~H.}\ \bibnamefont {Liew}},\ }\href
  {http://arxiv.org/abs/2003.09569} {\bibfield  {journal} {\bibinfo  {journal}
  {arXiv:2003.09569}\ } (\bibinfo {year} {2020})}\BibitemShut {NoStop}%
\bibitem [{\citenamefont {Killoran}\ \emph {et~al.}(2018)\citenamefont
  {Killoran}, \citenamefont {Bromley}, \citenamefont {Arrazola}, \citenamefont
  {Schuld}, \citenamefont {Quesada},\ and\ \citenamefont
  {Lloyd}}]{Killoran2018}%
  \BibitemOpen
  \bibfield  {author} {\bibinfo {author} {\bibfnamefont {N.}~\bibnamefont
  {Killoran}}, \bibinfo {author} {\bibfnamefont {T.~R.}\ \bibnamefont
  {Bromley}}, \bibinfo {author} {\bibfnamefont {J.~M.}\ \bibnamefont
  {Arrazola}}, \bibinfo {author} {\bibfnamefont {M.}~\bibnamefont {Schuld}},
  \bibinfo {author} {\bibfnamefont {N.}~\bibnamefont {Quesada}}, \ and\
  \bibinfo {author} {\bibfnamefont {S.}~\bibnamefont {Lloyd}},\ }\href@noop {}
  {\bibfield  {journal} {\bibinfo  {journal} {Physical Review Research}\
  }\textbf {\bibinfo {volume} {1}},\ \bibinfo {pages} {033063} (\bibinfo {year}
  {2018})}\BibitemShut {NoStop}%
\bibitem [{\citenamefont {Schuld}\ and\ \citenamefont
  {Killoran}(2019)}]{Schuld2019}%
  \BibitemOpen
  \bibfield  {author} {\bibinfo {author} {\bibfnamefont {M.}~\bibnamefont
  {Schuld}}\ and\ \bibinfo {author} {\bibfnamefont {N.}~\bibnamefont
  {Killoran}},\ }\href {\doibase 10.1103/PhysRevLett.122.040504} {\bibfield
  {journal} {\bibinfo  {journal} {Physical Review Letters}\ }\textbf {\bibinfo
  {volume} {122}},\ \bibinfo {pages} {40504} (\bibinfo {year}
  {2019})}\BibitemShut {NoStop}%
\bibitem [{\citenamefont {Govia}\ \emph {et~al.}(2020)\citenamefont {Govia},
  \citenamefont {Ribeill}, \citenamefont {Rowlands}, \citenamefont {Krovi},\
  and\ \citenamefont {Ohki}}]{Govia2020}%
  \BibitemOpen
  \bibfield  {author} {\bibinfo {author} {\bibfnamefont {L.~C.~G.}\
  \bibnamefont {Govia}}, \bibinfo {author} {\bibfnamefont {G.~J.}\ \bibnamefont
  {Ribeill}}, \bibinfo {author} {\bibfnamefont {G.~E.}\ \bibnamefont
  {Rowlands}}, \bibinfo {author} {\bibfnamefont {H.~K.}\ \bibnamefont {Krovi}},
  \ and\ \bibinfo {author} {\bibfnamefont {T.~A.}\ \bibnamefont {Ohki}},\
  }\href@noop {} {\bibfield  {journal} {\bibinfo  {journal} {arXiv:2004.14965}\
  } (\bibinfo {year} {2020})}\BibitemShut {NoStop}%
\end{thebibliography}
\end{document}